# Orbital Angular Momentum Locking via Bound States in the Continuum

Enrico Baù[4], Connor Heimig[1,4], Jonas Biechteler[1,4], Lina Rohrer[1,4], Michael Hirler[1,4], Haoran Ren[2], Stefan A. Maier[2,3], Alexander A. Antonov[1,4], and Andreas Tittl*[,1,4]

[1]Institute of Photonics, Hamburg University of Technology, 21073 Hamburg, Germany.

[2]School of Physics and Astronomy, Monash University, Clayton, Victoria 3800, Australia.

[3]Department of Physics, Imperial College London, London SW7 2AZ, United Kingdom.

[4]Chair in Hybrid Nanosystems, Nano-Institute Munich, Department of Physics, LMU Munich, Germany.

*Corresponding Author: andreas.tittl@tuhh.de

## Abstract

Optical vortices are electromagnetic fields twisting around a phase singularity, resulting in quantized orbital angular momentum (OAM). When such vortices are formed by evanescent hybrid light-matter quasiparticles known as polaritons, they are referred to as polaritonic vortices (PVs). The nanometer-scale topologically robust features of such PVs promise to enable applications for lasing and thermal emission at deeply subwavelength scales. However, many conventional techniques are prone to producing multimode PVs due to poor mode selectivity, resulting in OAM mixing that degrades vortex purity and limits their performance for high-fidelity optical information encoding and multi-dimensional imaging. To overcome this limitation, we introduce a platform that generates deeply subwavelength PVs through quasi-bound states in the continuum (qBICs) in dielectric metasurfaces. In contrast to existing approaches, the qBIC intrinsically locks the PV to a single OAM and makes it robust against the polarization state of the excitation, including linear, elliptical and circular polarization. We experimentally realize qBIC-driven PVs through the interference of hyperbolic phonon polaritons (HPhPs) in hexagonal boron nitride by exploiting the highly uniform out-of-plane electric fields generated by the photonic qBIC, characterized via scattering scanning near-field optical microscopy. This results in HPhPs with a wavelength of around 30-40 smaller than the incident light, thereby enabling ultra-dense packing of multiple robust PVs with distinct OAM. Our platform brings PVs to the photonic chip scale, enabling applications in structured optical information transfer and communications.

## Main

Optical vortices are electromagnetic fields which twist around a central axis, forming a helical phase structure with zero intensity at the center.[1] They carry orbital angular momentum (OAM)[2] with a well-defined topological charge $\ell$, which denotes how many times the phase winds by $2\pi$ around the vortex center. Optical vortices can manifest both in the far-field in free space beams and in the near-field within polaritonic systems. Initial implementations of far-field vortices have been realized through spatial light modulators[3,4] or arrays of nanoresonators[5–7] that imprint OAM onto the incident wavefront, with applications for imaging[8,9], metrology[10,11] and microrobotics.[12] However, their integration into compact photonic chips for optical communication and information processing is challenging[13,14] due to constraints of the diffraction limit, which imposes a fundamental ceiling on the achievable optical resolution. A promising solution to this challenge are near-field polaritonic vortices (PVs), which can be generated using highly confined surface waves, such as surface plasmon polaritons[15,16], surface phonon polaritons in isotropic materials[17] and hyperbolic phonon polaritons (HPhPs) in anisotropic materials[18,19]. This conversion of incident light to surface waves has been instrumental in generating PVs that possess spatial features far smaller than the free-space wavelength, with potential applications for nanophotonic thermal emitters[20], optical tweezing[21,22], nanoparticle rotation[23] as well as optical[24] and quantum[25,26] information processing and spin-orbit conversion.[27]

Commonly, PVs are excited by coupling circularly polarized light, which inherently carries spin angular momentum (SAM) associated with the temporal rotation of the electric field, to nanostructured polaritonic materials.[18,19] Such structures generally consist of enclosed Archimedean spirals with multiple arms to achieve the necessary phase accumulation for obtaining the desired OAM.[28,29] Thus, the incoming SAM is converted into OAM and an associated topological charge $\ell$, with contributions from both the vorticity $L$ imprinted by the spiral-shaped coupling structures and the SAM of the incident light $\sigma$ ($\ell = L + \sigma$). In practice, however, conventional approaches to excite such PVs often lead to the generation of multimode vortices with more than a single topological charge $\ell$, either due to imperfect polarization control, fabrication uncertainties or finite excitation laser linewidth. The resulting superposition reduces the purity of the generated vortices and limits their applicability towards mode-sensitive applications, such as high-capacity optical data transmission and communication.[16] While significant OAM selectivity has been achieved by exploiting strong nonlinear effects[30–32], such approaches significantly constrain the choice of materials and offer only limited control over the spectral window of selectivity.

Here, we show the generation of robust PVs with distinct OAM that are intrinsically created and locked via photonic quasi-bound states in the continuum (qBICs).[33–37] As opposed to conventional approaches, the OAM of our PVs is governed exclusively by the vorticity imposed by the structure ($\ell = L$). This is achieved by launching HPhPs[38,39] in hexagonal boron nitride (hBN) with linearly polarized light through qBICs, which emerge from periodic arrays of silicon high refractive index nanoresonators underneath the hBN with a broken in-plane symmetry within each unit cell. We experimentally characterize our qBIC-driven PVs via transmission-mode scattering scanning near-field optical microscopy (s-SNOM).[40–42] In contrast to previous approaches, our platform enables not only the locking of OAM to a single value of $\ell$ but also tunable control over the spectral selectivity window via the quality-factor ($Q$- factor) of the resonance, which can be readily adjusted through simple modifications of the metasurface geometry. In addition, our PVs are generated with a fixed $\ell$ independent of the polarization state (circular, elliptical, or linear) and remain robust against large variations in orientation angle of the polarization of up to ~70°. These capabilities may enable applications in all-optical logic networks and high-speed photonic chips.

**OAM selectivity via qBICs**

The resonators in our metasurface consist of enclosed Archimedean spirals with single or multiple arms (**Fig. 1a**). Specifically, alternating rows contain spirals of left and right handedness, enabling both positive and negative values of $\ell$ within each unit cell. By laterally offsetting each resonator pairwise, a qBIC emerges, with its linewidth controlled by the lateral offset $\Delta x$ (**Fig. S1**). The resulting out-of-plane electric fields are generated with a uniform phase across the entire resonator (**Fig. S2**, middle panel). A similar resonance mechanism has recently been applied for the generation of tunable polaritonic topologies.[43] When transferring hBN on top of these dielectric resonators and designing the structure such that the qBIC lies within the Reststrahlenband of hBN (**Supplementary Note 1**, **Fig. S3**), HPhPs are launched in the thin hBN layer with the same phase across the entire resonator edge through in-plane momentum transfer, forming a PV with integer $\ell$ (**Fig. S2**, right panel).

Since we excite with linearly polarized light which has no intrinsic SAM, the structural contribution $L$ is equal to the total topological charge $\ell$ and can be simply written as $L = n \cdot m$, where $n$ is the number of discrete spiral arms (each arm induces an effective geometric topological charge of ±1, where the sign depends on the direction of rotation of the spiral) and $m$ is an integer which describes a multiple of the HPhP wavelength in the distance between outer and inner radius of the spiral. For the remainder of this work, we set $m = 1$, since it provides the largest possible area on the resonator, thus reducing edge effects. In principle, $m$ can be arbitrarily high if a small enough HPhP wavelength $\lambda_{HPhP}$ is chosen. While the excitation of a single structure with linearly polarized light without the presence of the photonic qBIC produces multimode PVs, the qBIC locks the PV to a single OAM ($\ell = -1$) and all other contributions vanish (**Fig. 1b**). The purity of the resulting PVs can be quantified via an overlap integral[17] (**Eq. 1**) and is maximized when shifting the resonance to the excitation frequency of the PV, while all other contributions ($\ell \neq -1$) collapse to 0.

Unlike conventional approaches to generate PVs, our qBIC-driven PVs are robust against the incident polarization state (**Fig. 1c**), owing to the fact that the OAM of the PV originates solely from the qBIC. The ellipticity angle is defined as $\chi = \tan^{-1}(\frac{a}{b})$, where $a$ and $b$ are the ratio between major and minor axis of the polarization ellipse (**Fig. S4**). Our qBIC-driven PVs retain their topological charge $\ell = L$ for any $\chi$ even when exciting the metasurface with circular ($\chi = \pm 45°$) or elliptical ($0° < |\chi| < 45°$) instead of linear polarization ($\chi = 0°$) (**Fig. S4a**), as well as when varying the orientation angle of the polarization $|\Phi|$ up to around 70° (**Fig. S4b**). Additionally, the OAM locking behavior is compared for structures with two different offsets $\Delta x$, resulting in resonances with $Q \approx 500$ (**Fig. S5a**) and $Q \approx 100$ (**Fig. S5b**). Our simulations show that the width of the spectral window in which the OAM is fixed at a single value directly depends on $Q$, enabling precise geometric control over the locking behavior. A SEM image of a fabricated sample is shown in **Fig. 1d**, demonstrating excellent geometric reproduction and edge quality on each individual resonator, as well as showing the boundary between the structure without (blue-shaded) and with (orange-shaded) hBN on top.

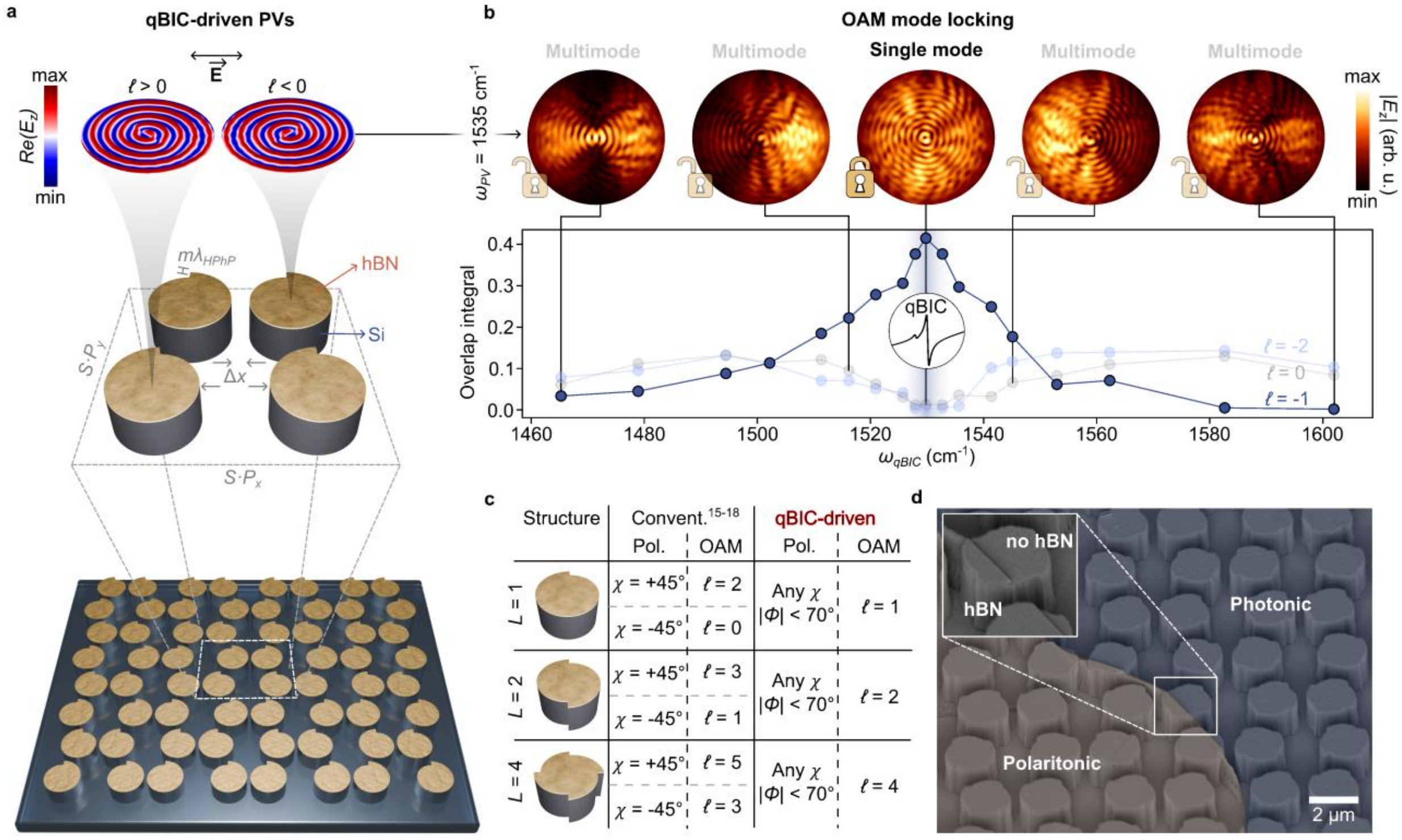

| Structure | | Convent.[15-18] Pol. | Convent.[15-18] OAM | qBIC-driven Pol. | qBIC-driven OAM |
|---|---|---|---|---|---|
| L = 1 | | χ = +45° | ℓ = 2 | Any χ \|Φ\| < 70° | ℓ = 1 |
| | | χ = -45° | ℓ = 0 | | |
| L = 2 | | χ = +45° | ℓ = 3 | Any χ \|Φ\| < 70° | ℓ = 2 |
| | | χ = -45° | ℓ = 1 | | |
| L = 4 | | χ = +45° | ℓ = 5 | Any χ \|Φ\| < 70° | ℓ = 4 |
| | | χ = -45° | ℓ = 3 | | |

**Figure 1: OAM locking of PVs via qBICs. a** Illustration of the PV-generating qBIC metasurface (bottom) consisting of hBN-covered silicon resonators on $CaF_2$ substrate that are fabricated in the shape of an enclosed Archimedean spiral. The unit cell (middle) consists of four resonators with in-plane pitches of $S \cdot P_x$, $S \cdot P_y$, where $S$ denotes the in-plane scaling factor. The distance between beginning and end of the spiral is a multiple of the polariton wavelength $m\lambda_{HPhP}$. The lateral offset between individual resonator pairs $\Delta x$ determines the asymmetry within each unit cell and thus the spectral linewidth of the resulting qBIC. The insets above the unit cell show analytically calculated real part of the out-of-plane electric fields $Re(E_z)$ for both right-handed ($\ell = -1$) and left-handed ($\ell = +1$) vortices. **b** Simulated overlap integrals (**Eq. 1**) of PVs generated from the structure shown in (**a**) ($\ell < 0$), excited at $\omega_{PV} = 1535\ \text{cm}^{-1}$ for varying qBIC spectral positions $\omega_{qBIC}$. Curves show calculated overlap integrals for $\ell = 0$ (light gray), $\ell = -1$ (dark blue) and $\ell = -2$ (light blue). The insets show the calculated amplitude of the out-of-plane electric field $|E_z|$ of selected PVs. Simulations were conducted for $\omega_{PV} = 50$ nm and pitches ranging from $P_x = 5150 - 5950$ nm, $P_y = 4635 - 5355$ nm. The resonance of the simulated structure has a quality factor of around $Q \approx 500$. $m\lambda_{HPhP}$ with m = 1 is kept constant for all simulations. Inset shows a simulated reflectance spectrum of a qBIC around 1535 $\text{cm}^{-1}$. **c** Comparison between polarization of excitation and topological charges $\ell$ emerging from conventional structures (left) requiring circular polarization ($\chi = \pm 45°$), and from the metasurface used in our work (right), which works for any polarization state/ellipticity angle $\chi$ and for $|\Phi| < 70°$. **d** SEM image of the fabricated metasurface for the generation of PVs with $\ell = \pm 4$. The blue and orange shaded areas show the photonic (no hBN) and polaritonic (hBN-covered) areas respectively.

## Origin of OAM-generating qBIC

To construct the photonic qBIC that is used to generate the PVs, we start from a simple square lattice consisting of subwavelength dielectric disks, with each unit cell containing one resonator (**Fig. 2a**, left panel). This structure supports a non-radiating ($\gamma_{rad} = 0$), non-degenerate symmetry-protected BIC[33,34] with in-plane magnetic field and, consequently, an out-of-plane electric field $E_z$ component (**Fig. S6**). To access the qBIC, it is necessary to perform both Brillouin zone folding[44,45] and break the rotational symmetry. We achieve both simultaneously by introducing a lateral offset $2\Delta x$ for every second resonator in a checkerboard pattern (**Fig. 2a**, middle panel). The factor of 2 is introduced here to keep $\Delta x$ consistent with **Fig. 1a**. This results in the lattice period of the metasurface increasing by a factor of $\sqrt{2}$. Through this perturbation, the qBIC acquires a radiative loss channel ($\gamma_{rad} \neq 0$) and a finite $Q$-factor determined by the asymmetry $\Delta x$ (**Fig. S1c**). Note that the unit cell shown earlier in **Fig. 1a** is not the smallest possible unit cell, being larger by an additional factor $\sqrt{2}$.

We explore the far-field characteristics of the qBIC for the structure shown in the middle panel of **Fig. 2a**. We simulate both the resonance position (**Fig. 2b**) and $Q$-factor (**Fig. 2c**) in momentum space, assuming a fixed permittivity of $\varepsilon = 10.6$ and no substrate. The radius and height of each resonator is set to 1000 nm and 1450 nm, respectively; the offset of the central resonator is 450 nm, and the period (in accordance with **Fig.1a**) is defined as $P_x = P_y = 5000$ nm. Both the resonance position and the $Q$-factor remain nearly constant over a relatively broad region of momentum space, exhibiting weak negative and positive parabolic dispersions, respectively. Interestingly, the far-field polarization distribution of the qBIC carries no topological charge. Thus, for all angles of incidence, the far-field signal is polarized along the direction of the displacement $\Delta x$ (**Fig. 2d**). This behavior provides the robustness of our PVs against changes in polarization of the excitation (**Fig. S4**). Next, each disk is transformed into an Archimedean spiral to encode OAM (**Fig. 2a**, right panel). This perturbs the photonic mode only weakly due to the distance between outer and inner radius of the spiral matching the HPhP wavelength, which is generally much smaller than both the incident wavelength and unit cell size (by a factor of 20-30 in our case, see **Fig. S7**). Furthermore, our design enables the generation of PVs with both positive and negative OAM within each unit cell by alternating the handedness of the Archimedean spirals from row to row, thus offering a straightforward way to achieve OAM multiplexing.

To generate PVs with arbitrary OAM, a matching number $n$ of arms can be added to the resonator (**Fig. 2e**). Since we only consider Archimedian spirals with distances between outer and inner radii that exactly match a single polariton wavelength $\lambda_{HPhP}$ ($m = 1$), to maximize the area available on the resonator surface and avoid edge effects, the total OAM $\ell$ equals $n$ in our case. The dominant out-of-plane electric fields generated by the photonic qBIC then launch HPhPs with the same amplitude and phase along the entire spiral edge through in-plane momentum transfer (**Fig. 2f**). The fact that these vortices can be observed on the surface of subwavelength resonators is owed to the deeply subwavelength dispersion of hBN in the mid-IR region of light as well as the material's intrinsic anisotropy, which enables the generation of HPhPs with in-plane momenta that are tens or hundreds of times larger than the momentum of the incident light (**Fig. 2g**).

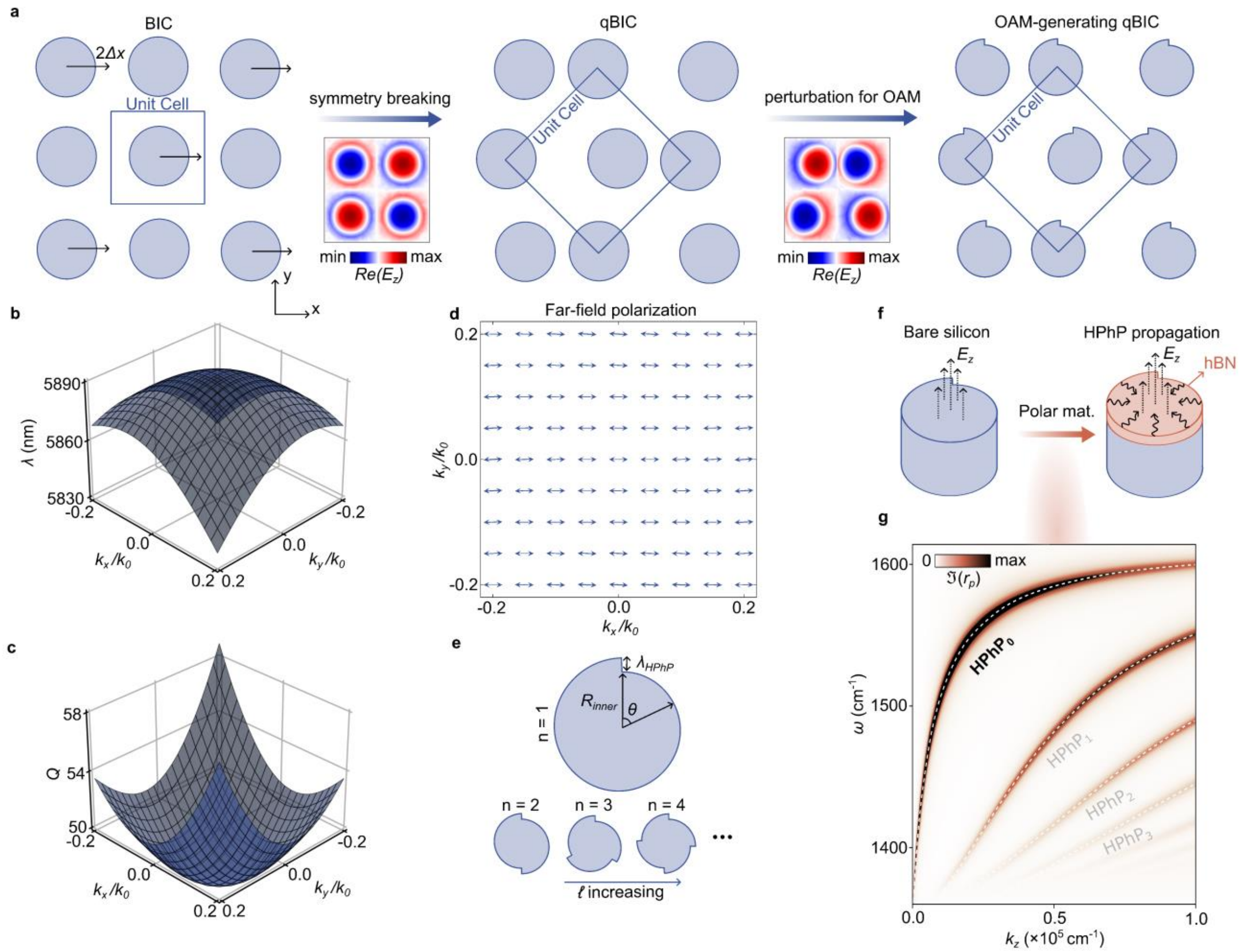

**Figure 2. Origin and working principle of OAM-generating qBIC. a** A periodic lattice of dielectric disks has a BIC (left panel) which can be accessed from the far-field by periodically shifting every other resonator in a checkerboard pattern, thus resulting in a qBIC with a finite resonance linewidth (middle panel). By transforming each disk into an Archimedean spiral, the structure supports deeply subwavelength PVs when coupled to hBN without perturbing the photonic mode significantly. Hereby, the spirals alternate in handedness from row to row, enabling the generation of PVs with both positive and negative OAM within each unit cell. The insets show the real part of the out-of-plane electric field $E_z$ of the unperturbed BIC (left) and qBIC (right). **b** Resonance position, **c** $Q$-factor and **d** far-field polarization in momentum space of the photonic qBIC of the structure shown in the middle panel of (**a**). **e** Archimedean spiral design with different numbers of arms $n$, azimuthal angle $\theta$ and distance between outer and inner spiral radius equaling to $\lambda_{\mathrm{HPhP}}$. $\mathrm{R_{inner}}$ denotes the inner radius of an individual spiral resonator**. f** Schematic representation of the out-of-plane electric fields $E_z$ generated by the photonic qBIC that launch HPhPs in the hBN through in-plane momentum transfer with the same amplitude and phase from each edge. **g** Calculated imaginary part of the reflection coefficient $\Im(r_p)$ for various $\omega$ and HPhP momenta $k_z$, showing the various hyperbolic polariton branches $\mathrm{HPhP_j}$. The dielectric function of hBN has been modeled according to Ref.[39] (**Fig. S3**).

## Generation of high-purity PVs

The physical properties of our qBIC-driven vortices are studied through simulations of PVs with various OAM ranging from $\ell = 1$ to $\ell = 11$ (**Fig. 3**). For all simulations, an asymmetry of $\Delta x = 0.045P_x$ and a flake thickness of $h_{hBN} = 50$ nm were chosen to ensure sufficiently high $Q$-factor and long HPhP lifetimes. The resulting reflectance spectra (**Fig. 3b**) consistently show modulation above 90% and $Q$-factors between 110 and 130 irrespective of $n$. We chose structures supporting modest $Q$-factors to compare simulations with experiments shown in the following section. In **Fig. 3c**, we show the simulated out-of-plane near-field phase of various right-handed PVs at a wavenumber of around 1550 cm$^{-1}$, matching the resonances shown in **Fig. 3b**. To remove the intrinsic material contributions from hBN that manifest as low-frequency components, we Fourier-filter the simulated data in the same way as the experimental data shown later (**Fig. S8**), similar to previous works.[17,46] With an excitation wavelength of $\lambda_0 = 6500$ nm, the resulting PVs are generated at a deeply subwavelength scale with $\frac{\lambda_{HPhP}}{\lambda_0} = \frac{1}{36}$, with polariton wavelengths much smaller than conventional plasmonic vortices ($\frac{\lambda_{Plasmon}}{\lambda_0} \approx 1$).

To evaluate the purity of our qBIC-driven PVs, we calculate an overlap integral $OI_\ell$ between measured and analytically calculated PVs. For a PV with a given momentum $\ell$, this can be written as:

$$OI_\ell = \frac{|\int E_z \bar{E}^*_{z,\ell} dr d\theta|^2}{\int E_z E^*_z dr d\theta \int \bar{E}_{z,\ell} \bar{E}^*_{z,\ell} dr d\theta} \quad (1)$$

where $E_z$ is the simulated or experimentally measured complex-valued out-of-plane electric field, $\bar{E}_{z,\ell}$ are analytically calculated field distributions and $E^*_z$ and $\bar{E}^*_{z,\ell}$ their conjugate counterparts. The exact procedure to calculate $\bar{E}_{z,\ell}$ has been described in previous works.[17,47] Integration is performed within the cross-section of a single resonator 5 nm above the hBN layer, where $\theta$ is the azimuthal angle and $r$ is the distance from the central point of the resonator of any given point. First, we calculate $OI_\ell$ based on the simulated complex-valued $E_z$. Our qBIC-driven PVs exhibit high purity ($OI_\ell > 0.4$) (**Fig. 3d**), with the main contributor stemming from $\ell = n$.

The high purity obtained for the largest simulated value of $\ell = 11$ indicates that PVs carrying large OAM should in principle be achievable with our platform. Since the photonic mode is largely unaffected by the number of arms within each resonator, we estimate that a reasonable limit for the maximum achievable OAM using our platform, assuming $m = 1$, equals approximately $\ell_{max} < \frac{2\pi R_{inner}}{\lambda_{HPhP}}$, where $R_{inner}$ denotes the inner radius of an individual resonator. The reasoning behind this approximation comes from the fact that, if $\lambda_{HPhP}$ is smaller than the length of a single spiral arm, both the launching efficiency and phase accumulation will likely be insufficient, leading to distorted near-field patterns and thus low PV purity. Since our simulated structures have an inner radius of around $R_{inner} = 1.1$ µm and support HPhPs with $\lambda_{HPhP} = 180$ nm, this yields an estimated maximum value of $\ell_{max} = 38$. Note that the polariton dispersion (**Fig. 2g**) can be exploited to generate HPhPs with much smaller wavelength, either by shifting the resonance to higher wavenumbers or choosing thinner flakes to increase the in-plane momentum of HPhPs. Since HPhPs are commonly observed to be in the range of 100 nm, this could easily enable PVs with $\ell_{max} > 50$ or even higher OAM.

To further demonstrate the capabilities of our platform, we simulate a multiplexed unit cell comprising four resonators, each with a different number of arms $n$ and therefore capable of simultaneously generating four spatially separated PVs with distinct values of $\ell = 1 - 4$ (**Fig. 3e**). Despite the increase in perturbation arising from the varying resonator geometries within each unit cell, both the reflectance modulation and the $Q$-factor of the qBIC remain relatively high (**Fig. 3f**). Specifically, the reflectance modulation decreases by around 20%, while the $Q$-factor reduces by around 10-20% compared to the cases shown in **Figs. 3a–d**. The calculated overlap integrals show that each resonator generates a PV with exactly $\ell = n$ (**Fig. 3g**). Notably, the purity of each PV remains consistently high (above 0.4), enabling the high-density integration of four distinct PVs within a single subwavelength unit cell.

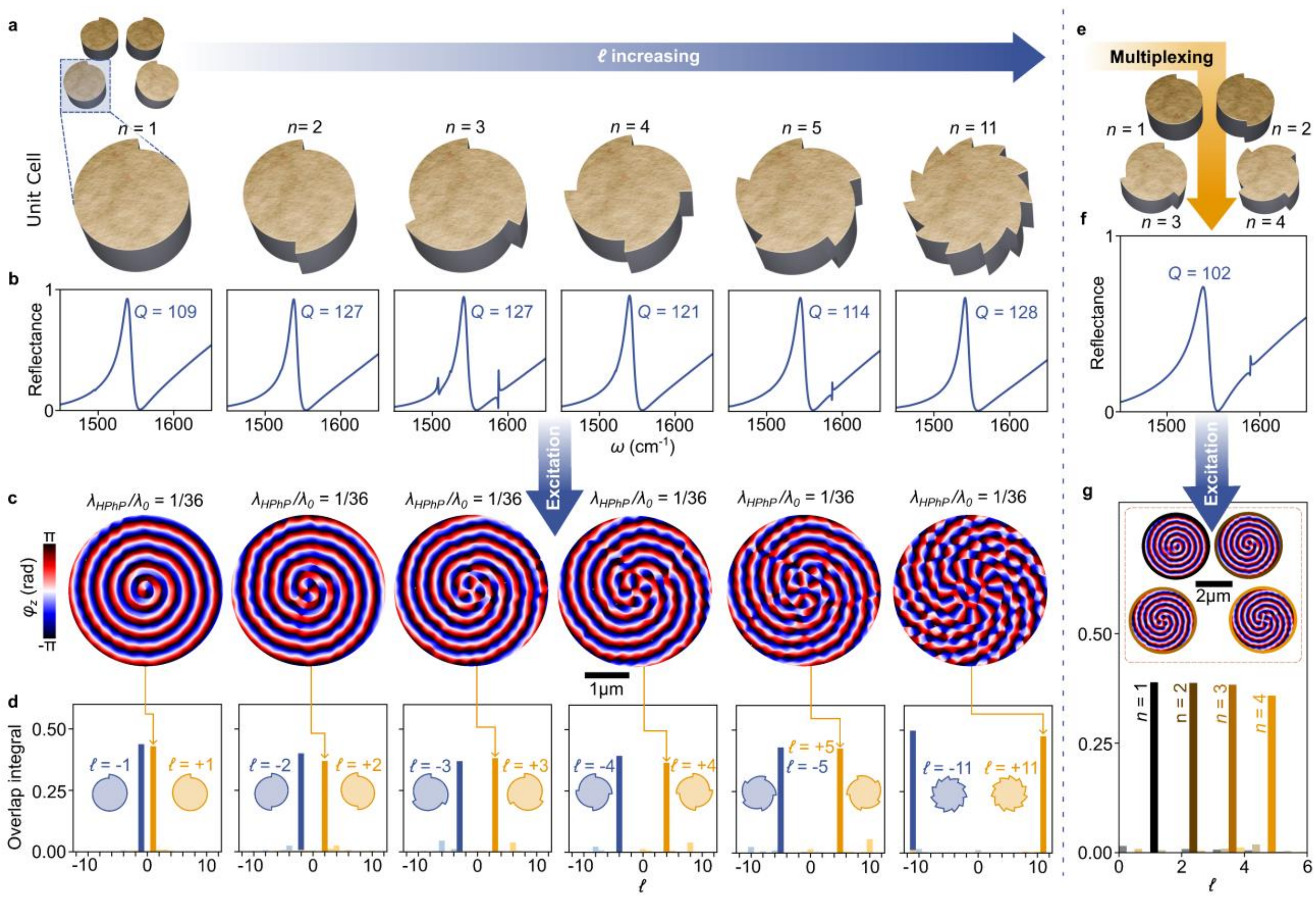

**Figure 3**. **Generation of high-purity qBIC-driven PVs with arbitrary** $\ell$**. a** Illustration of single unit cells/resonators for the generation of PVs with arbitrary $\ell$. **b** Simulated reflectance spectra and extracted $Q$-factors of the structures shown in (**a**). **c** Simulated near-field optical phase $\varphi_z$ of PVs with various $\ell$ between 1 and 11. Only right-handed PVs are shown. **d** Overlap integrals $OI_\ell$ calculated from d for left-handed (blue) and right-handed (orange) PVs with reference functions $\bar{E}_{z,\ell\in[-11,11]}$. **e** Illustration of a multiplexed unit cell consisting of four resonators ranging from $n = 1-4$. **f** Simulated reflectance spectra and extracted $Q$-factor of the unit cell shown in (**e**). **g** Simulated near-field optical phase $\varphi_z$ and calculated overlap integrals $OI_\ell$ of the multiplexed unit cell shown in (**e**).

We begin our experimental investigation by measuring the reflectance of the fabricated metasurfaces. The pitch for an in-plane scaling factor of $S = 1$ is set to $P_x = 5000$ nm and $P_y = 4500$ nm (see **Fig. 1a**) and the height of the silicon is to $h_{Si} = 1450$ nm. The asymmetry is set to $\Delta x/P_x = 0.045$ and the inner radius of the spiral to around $R_{inner} = 1.1$ μm. As our design incorporates spirals of both handednesses, it enables the generation of counter-rotating vortices with opposite $\ell$ within each unit cell. For all arrays, the distance between outer and inner radius of the spiral is set to around 180 nm, matching the wavelength of HPhPs launched at the resonance position. To ensure that this precise condition is met and to account for slight fabrication uncertainties, the metasurface is fabricated as a spectral gradient to cover a range of wavelengths. This is achieved by smoothly varying the in-plane scaling factor $S$ across the metasurface.[48–50] For a sketch of the fabrication procedure, see **Fig. S9**. An optical image (**Fig. S10a**) and a SEM image (**Fig. S10b**) show the fabricated array consisting of resonators that generate vortices with $\ell = 1$.

Reflectance far-field spectra of the spectral gradient metasurface are acquired across the Reststrahlenband region of hBN (see **Materials and Methods**). The measurements (**Fig. S10c**) demonstrate a qBIC with a quality factor of around $Q = 50 - 100$ for structures of various scaling factors. A redshift of the resonance can be observed for higher scaling factors, consistent with numerically simulated results (**Fig. S10d**). Compared to conventional qBIC metasurfaces, our structures are intended to support moderate $Q$-factors in order to avoid full quenching of the qBIC due to intrinsic losses within the hBN Reststrahlenband. Using the spectral positions of the resonances determined through our far-field measurements, we characterize both the near-field structure and purity of the resulting PVs.

We experimentally demonstrate the generation of qBIC-driven PVs via transmission-mode s-SNOM.[40–42] This enables us to measure the near-field optical amplitude $|E_z|$ as well as the phase $\varphi_z$ of the out-of-plane electric field of deeply subwavelength PVs induced by the qBIC mode. This method has been well established for the characterization of topological textures of light, such as optical skyrmions[46,51], as well as for probing the near-field of dielectric metasurfaces.[52] Since we utilize pseudo-heterodyne interferometry (see **Materials and Methods**) and the tip shaft is mostly polarized in the out-of-plane direction, both near-field optical amplitude $|E_z|$ and phase $\varphi_z$ of the out-of-plane electric field can be retrieved simultaneously. A schematic of the setup is shown in **Fig. S11**. The near-field measurements of $|E_z|$ and $\varphi_z$ and the (simultaneously) measured topography, are shown in **Fig. 4** for structures supporting $\ell = 1$ (**Fig. 4a)**, $\ell = 2$ (**Fig. 4b)** and $\ell = 4$ (**Fig. 4c)**. Similar to the simulations shown in **Fig. 3**, the unprocessed near-field data (**Fig. S12**) is Fourier-filtered to remove low and high frequency components that can appear due to defects in the hBN or scanning artifacts (**Fig. S8**). Far-field reflectance spectra measured on the same metasurface as the near-field images exhibit high modulation (> 0.7) and modest $Q$-factors between 61 and 74.

We observe that, when comparing the polariton wavelength to the wavelength of the incident beam $\lambda_0$, the PVs generated by our qBIC metasurface are deeply subwavelength ($\frac{\lambda_{HPhP}}{\lambda_0} < \frac{1}{30}$) for all $\ell$, constituting a direct advantage of our near-field vortices over their free-space counterparts. Importantly, despite the metasurface being excited with linear polarization along the displacement direction, the filtered near-field images do not show significant distortions in the direction of polarization and most of the local intensity variations in the optical amplitude likely results from asymmetric near-field resonator-to-resonator interactions. As can be seen from each calculated overlap integral, the measured PVs have high purity ($OI_\ell > 0.3$), agreeing well with simulations shown in **Fig. 3** and are competitive with conventional PV-platforms[17]. This mode purity could likely be further improved by either fabricating larger resonators and thus minimizing edge effects or shrinking the HPhP wavelength through dispersive tuning. As seen from the charts, many of the side contributions stem from topological charges of $\ell = 0$ or from $\ell$ of opposite sign. These additional contributions may arise due to weak deuterogenic effects, which result in the coexistence of multiple topological charges and usually appear in higher-order PVs.[53]

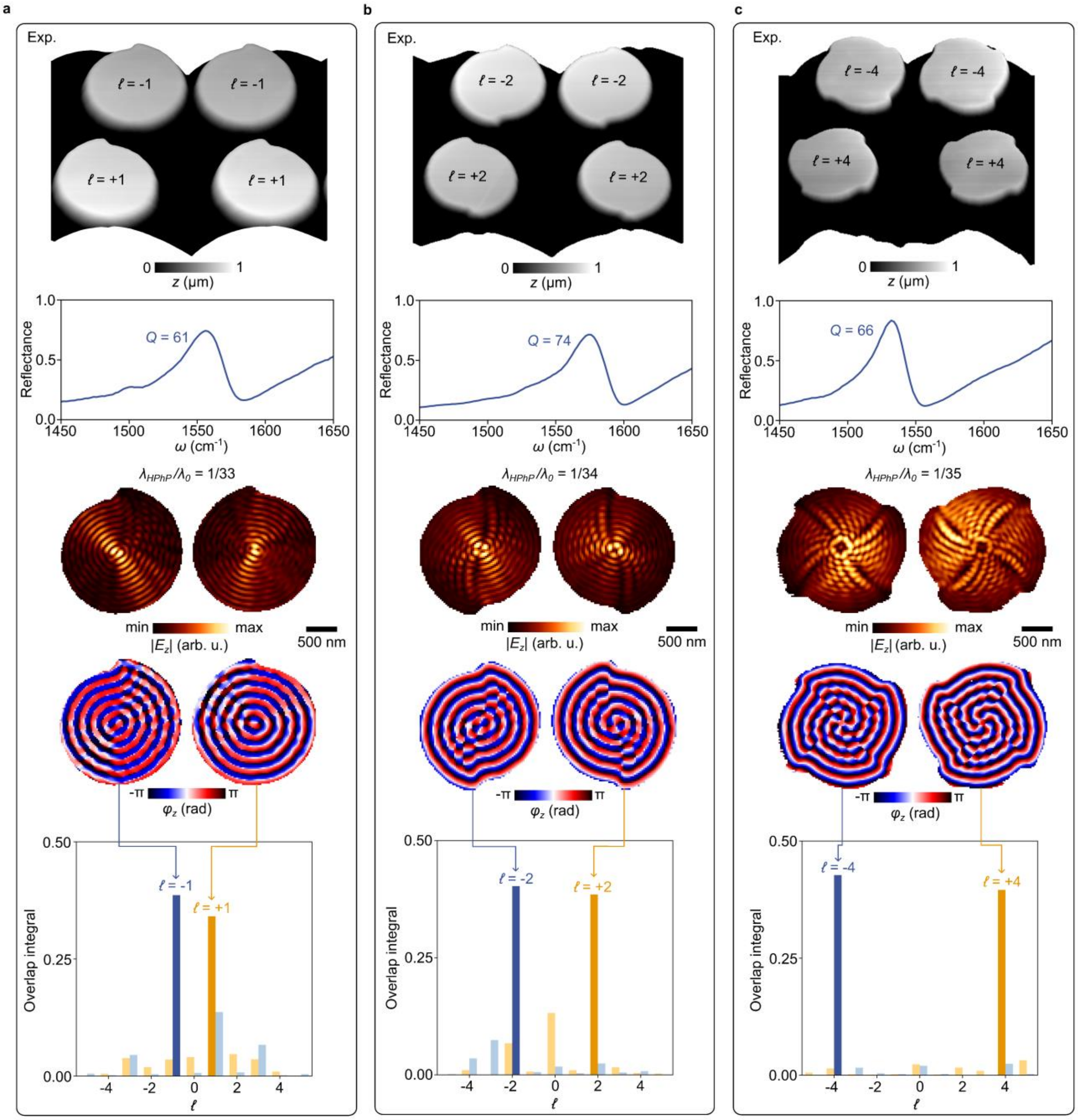

**Figure 4. Experimentally measured PVs with various OAM charges $\ell$. a** Experimental results for unit cell that supports $\ell = \pm 1$ (one spiral arm). From top to bottom: measured topography of a single unit cell, reflectance spectra with fitted $Q$-factor, near-field out-of-plane optical amplitude $|E_z|$ and phase $\varphi_z$ for both right-handed (left) and left-handed (right) PVs and calculated overlap integrals with reference functions $\bar{E}_{z,\ell\in[-5,5]}$. Same for $\ell = \pm 2$ (**b**) and $\ell = \pm 4$ (**c**). Measurements were taken on hBN flakes with thicknesses $h_{hBN}$ between 50 nm and 100 nm and excitation frequencies of 1555 cm$^{-1}$ (**a**), 1575 cm$^{-1}$ (**b**) and 1532 cm$^{-1}$ (**c**).

## Conclusion

We successfully demonstrated the generation of robust, deeply subwavelength ($\frac{\lambda_{HPhP}}{\lambda_0} < \frac{1}{30}$) PVs driven via qBICs. To achieve significant OAM selectivity, previous approaches had to rely on strong nonlinear material reponses.[30–32]

In contrast, the OAM of our PVs originates from the underlying photonic mode under linear excitation and is locked over a spectral range governed by the $Q$-factor of the resonance. Furthermore, our PVs maintain high modal purity for arbitrary polarization states and under large variations of the orientation angle of the polarization, enhancing the robustness of PVs against imperfect illumination conditions and making them suitable for practical technological applications such as single-mode lasers and highly localized information encoding. In addition, moving away from singular structures towards non-local resonances and large array-based photonic modes enables high-density packing of optical vortices within a single array that is broadly illuminated without the need for structured excitation, thus facilitating on-chip multiplexing.

We experimentally demonstrated PVs with OAM of $\ell = 1, 2, 4$ via near-field measurements with s-SNOM. Calculated overlap integrals show excellent agreement between theory and experiments, resulting in vortices that exhibit purity comparable to previous platforms.[17] In principle, our approach can be straightforwardly extended to obtain higher-order PVs by simply increasing the number of arms within each resonator. As we showed in simulations, this would not perturb the photonic mode in any significant way, since the offsets used to imprint the OAM are relatively small (< 200 nm) compared to the size of a single resonator (diameter of around 2-3 μm). Generating such a wide range of OAM within a relatively small spatial footprint is a unique capability of phonon-polaritonic vortices, as opposed to plasmonic vortices, where the plasmon wavelength in the low-loss regime remains approximately equal to the incident wavelength. Based on our estimation discussed earlier, our platform may support PVs with $\ell_{max} > 50$ or even higher OAM, satisfying requirements for current technological applications such as optical communication.[54]

Our approach could enable gradient metasurfaces in which the number of spiral arms varies spatially, allowing on-chip encoding of OAM. Furthermore, previous works have shown that optical topological lattices can exert optical forces on subwavelength particles, generating periodic trapping wells.[55] The trapping and control of the motion of nanoparticles or biomolecules below the wavelength of the HPhPs measured in this work (around 200 nm) could be of great interest for e.g. microrobotics[56] and biophysics by selectively collecting and trapping small cells or bacteria on the resonator surface, then conducting biosensing experiments based on a pixelated sensor approach[57] with spatially encoded OAM. Overall, combining the physics of photonic modes such as qBICs with PVs could be useful for engineering single-mode lasers[58], enhancing and structuring thermal emission[20,59,60], as well as exploiting and enhancing nonlinear effects[61].

## Materials and Methods

### Numerical Simulations

Characteristics of the photonic qBIC in momentum space were numerically investigated using the Electromagnetic Waves Frequency Domain module of COMSOL Multiphysics in 3D mode using a previously developed approach.[62] The tetrahedral spatial mesh for finite element method was automatically generated by COMSOL's physics-controlled preset. Simulations were performed within a rectangular spatial domain containing a single metasurface unit cell shown in the middle panel of **Fig. 2a** with periodic boundary conditions applied to its sides. Far-field spectra and near-fields of the PVs shown in this work were conducted using CST Studio Suite (Simulia), a commercial finite element solver. The setup includes adaptive mesh refinement and periodic boundary conditions in $x, y$ direction and open boundaries in $z$ direction. All simulations are conducted in the frequency domain. The numerically calculated near-fields were extracted 5 nm above the surface of the hBN layer to match the experimental s-SNOM measurements. The electric fields were then exported with 5 nm resolution.

**Fabrication**

The OAM-generating qBIC metasurfaces were realized through a multilayer fabrication process. First, an amorphous silicon film was deposited onto a $CaF_2$ substrate by plasma-enhanced chemical vapor deposition using a PlasmaPro 100 reactor (Oxford Instruments). A conformal $SiO_2$ coating was subsequently added via radio frequency sputtering in an Amod physical vapor deposition system (Angstrom Engineering). Exfoliated hBN flakes, obtained from bulk crystals (HQ Graphene), were transferred onto the substrate held at 160 °C to ensure good adhesion and removal of residual. Any remaining residues from the transfer process were removed by a 10-minute oxygen plasma cleaning step. Candidate flakes with suitable lateral dimensions and thickness uniformity were identified under an optical microscope and further characterized with a Bruker Dektak XT profilometer. The multilayer stack was then completed by depositing a chromium (Cr) layer in the aforementioned sputtering tool. For lithographic patterning, the sample was spin-coated with the positive-tone resist CSAR 62 (Allresist) and written on a Raith eLINE Plus electron beam lithography system operated at 20 kV with a 15 µm aperture. Development proceeded in two stages: an amyl acetate bath followed by immersion in a 1:9 MIBK:IPA solution. Reactive ion etching (RIE) via the PlasmaPro 100 (Oxford Instruments) was used to sequentially transfer the pattern through the stack: first Cr (serving as a hard mask after resist removal), then hBN, $SiO_2$, silicon, and finally Cr again to remove the remaining mask. A schematic illustration of the full process is provided in **Fig. S6**.

**Optical Far-field Measurements**

Far-field reflectance spectra were taken using a Spero mid-infrared spectral imaging microscope (Daylight Solutions Inc.). A 4× magnification objective ($NA = 0.15$) was used for all measurements, which provided a 2 × 2 $mm^2$ field of view with a resolution of 480 × 480 pixels and a pixel size of 4 × 4 $\mu m^2$. Three linearly-polarized quantum cascade lasers which can be tuned from 5.6 to 10.5 µm with a spectral resolution of 2 $cm^{-1}$ were used, thus covering the entirety of the in-plane hBN Reststrahlen-band.

**Optical Near-field Measurements**

The optical near-field measurements in this work were conducted with a commercial s-SNOM system (neaSCOPE from attocube systems, Haar, Germany). The laser source was a continuous wave quantum cascade laser (MirCat, Daylight Solutions), tunable from 1755 $cm^{-1}$ to 1315 $cm^{-1}$. The atomic force microscopy tip is operated in tapping mode, oscillating at a frequency of around 250 kHz. For all experiments, we used metal-coated (Platinum/Iridium) AFM tips (Arrow-NCPt, NanoWorld). S-SNOM enables the extraction of both out-of-plane near-field scattering amplitude $|E_z|$ and out-of-plane phase $\varphi_z$ via pseudo heterodyne detection.[63] To avoid artefacts in the images that can emerge due to the angle of incident typically seen in s-SNOM experiments in reflection mode, our measurements were done in transmission mode with a 0 degree angle of incidence to excite the photonic mode. Our setup contains a beam splitter which splits the incident beam into two parts, with the first hitting a parabolic mirror and being loosely focused onto the metasurface and the probing AFM tip from below. A second parabolic mirror collects the light backscattered from the tip. This backscattered beam then interferes with a reference beam, which passed through a delay stage containing two vibrating mirrors to decouple optical near-field amplitude and phase. The recombined beams are measured at a liquid nitrogen cooled mercury cadmium telluride detector. To suppress far-field contributions, we consider only higher harmonics of the signal demodulated at the tip frequency (**Fig. S11**).

**Acknowledgments**

This project was funded by the Deutsche Forschungsgemeinschaft (DFG, German Research Foundation) under grant numbers EXC 2089/1–390776260 (Germany's Excellence Strategy) and TI 1063/1 (Emmy Noether Program), the Bavarian program Solar Energies Go Hybrid (SolTech), and Enabling Quantum Communication and Imaging Applications (EQAP), and the Center for NanoScience (CeNS). H. R. acknowledges funding support from the Australian Research Council grants (DE220101085, DP220102152, FT250100565). Funded by the European Union (ERC, METANEXT, 101078018 and EIC, NEHO, 101046329). Views and opinions expressed are however those of the author(s) only and do not necessarily reflect those of the European Union or the European Research Council Executive Agency. Neither the European Union nor the granting authority can be held responsible for them. S.A.M. additionally acknowledges the Lee-Lucas Chair in Physics.

**Author contributions**

E.B. and A.T. conceived the idea and planned the research. C.H. and J.B. contributed to the sample fabrication. E.B. and L.R. performed the measurements. E.B. and A.A.A. conducted the numerical simulations. E.B., C.H., J.B., L.R., M.H., H.R. and A.A.A. contributed to the data processing. E.B., C.H., J.B., L.R., M.H., H.R., S.A.M., A.A.A. and A.T. contributed to the data analysis. S.A.M. and A.T. supervised the project. All authors contributed to the writing of the paper.

**Competing interests**

The authors declare no competing interest.

**Data and materials availability**

All data are available in the main text or in the supplementary materials.

**References**

1. Allen, L., Beijersbergen, M. W., Spreeuw, R. J. & Woerdman, J. P. Orbital angular momentum of light and the transformation of Laguerre-Gaussian laser modes. *Physical review. A, Atomic, molecular, and optical physics* **45,** 8185–8189; 10.1103/PhysRevA.45.8185 (1992).

2. Cardano, F. & Marrucci, L. Spin–orbit photonics. *Nature Photon* **9,** 776–778; 10.1038/nphoton.2015.232 (2015).

3. Wang, J. *et al.* Terabit free-space data transmission employing orbital angular momentum multiplexing. *Nature Photon* **6,** 488–496; 10.1038/nphoton.2012.138 (2012).

4. Anaya Carvajal, N., Acevedo, C. H. & Torres Moreno, Y. Generation of Perfect Optical Vortices by Using a Transmission Liquid Crystal Spatial Light Modulator. *International Journal of Optics* **2017,** 1–10; 10.1155/2017/6852019 (2017).

5. Devlin, R. C., Ambrosio, A., Rubin, N. A., Mueller, J. P. B. & Capasso, F. Arbitrary spin-to-orbital angular momentum conversion of light. *Science (New York, N.Y.)* **358,** 896–901; 10.1126/science.aao5392 (2017).

6. Ren, H. & Maier, S. A. Nanophotonic Materials for Twisted-Light Manipulation. *Advanced materials (Deerfield Beach, Fla.)* **35,** e2106692; 10.1002/adma.202106692 (2023).

7. Ahmed, H. *et al.* Optical metasurfaces for generating and manipulating optical vortex beams. *Nanophotonics (Berlin, Germany)* **11,** 941–956; 10.1515/nanoph-2021-0746 (2022).

8. Fürhapter, S., Jesacher, A., Bernet, S. & Ritsch-Marte, M. Spiral phase contrast imaging in microscopy. *Optics express* **13,** 689–694; 10.1364/OPEX.13.000689 (2005).

9. Li, L. & Li, F. Beating the Rayleigh limit: orbital-angular-momentum-based super-resolution diffraction tomography. *Physical review. E, Statistical, nonlinear, and soft matter physics* **88,** 33205; 10.1103/PhysRevE.88.033205 (2013).

10. Litchinitser, N. M. Applied physics. Structured light meets structured matter. *Science (New York, N.Y.)* **337,** 1054–1055; 10.1126/science.1226204 (2012).

11. Lavery, M. P. J., Speirits, F. C., Barnett, S. M. & Padgett, M. J. Detection of a spinning object using light's orbital angular momentum. *Science (New York, N.Y.)* **341,** 537–540; 10.1126/science.1239936 (2013).

12. Wu, L. *et al.* Micromotor based on single fiber optical vortex tweezer. *APL Photonics* **9**; 10.1063/5.0232282 (2024).

13. Yue, Z., Ren, H., Wei, S., Lin, J. & Gu, M. Angular-momentum nanometrology in an ultrathin plasmonic topological insulator film. *Nature communications* **9,** 4413; 10.1038/s41467-018-06952-1 (2018).

14. Ji, Z. *et al.* Photocurrent detection of the orbital angular momentum of light. *Science (New York, N.Y.)* **368,** 763–767; 10.1126/science.aba9192 (2020).

15. Spektor, G. *et al.* Orbital angular momentum multiplication in plasmonic vortex cavities. *Science advances* **7**; 10.1126/sciadv.abg5571 (2021).

16. Ren, H., Li, X., Zhang, Q. & Gu, M. On-chip noninterference angular momentum multiplexing of broadband light. *Science (New York, N.Y.)* **352,** 805–809; 10.1126/science.aaf1112 (2016).

17. Mancini, A. *et al.* Multiplication of the orbital angular momentum of phonon polaritons via sublinear dispersion. *Nature Photon* **18,** 677–684; 10.1038/s41566-024-01410-5 (2024).

18. Wang, M. *et al.* Spin-orbit-locked hyperbolic polariton vortices carrying reconfigurable topological charges. *eLight* **2**; 10.1186/s43593-022-00018-y (2022).

19. Xiong, L. *et al.* Polaritonic Vortices with a Half-Integer Charge. *Nano letters* **21,** 9256–9261; 10.1021/acs.nanolett.1c03175 (2021).

20. Baranov, D. G. *et al.* Nanophotonic engineering of far-field thermal emitters. *Nature materials* **18,** 920–930; 10.1038/s41563-019-0363-y (2019).

21. Shen, Z. & Liu, N. Optical Tweezers with Optical Vortex Based on Deep Learning. *ACS Photonics* **12,** 2212–2218; 10.1021/acsphotonics.5c00137 (2025).

22. Zhang, Y. *et al.* Plasmonic tweezers: for nanoscale optical trapping and beyond. *Light, science & applications* **10,** 59; 10.1038/s41377-021-00474-0 (2021).

23. Shen, Z. *et al.* Visualizing orbital angular momentum of plasmonic vortices. *Optics letters* **37,** 4627–4629; 10.1364/OL.37.004627 (2012).

24. Yan, Y. *et al.* High-capacity millimetre-wave communications with orbital angular momentum multiplexing. *Nature communications* **5,** 4876; 10.1038/ncomms5876 (2014).

25. Nagali, E. *et al.* Quantum information transfer from spin to orbital angular momentum of photons. *Physical review letters* **103,** 13601; 10.1103/PhysRevLett.103.013601 (2009).

26. Fickler, R. *et al.* Interface between path and orbital angular momentum entanglement for high-dimensional photonic quantum information. *Nature communications* **5,** 4502; 10.1038/ncomms5502 (2014).

27. Kotlyar, V. V., Nalimov, A. G., Kovalev, A. A., Porfirev, A. P. & Stafeev, S. S. Spin-orbit and orbit-spin conversion in the sharp focus of laser light: Theory and experiment. *Phys. Rev. A* **102**; 10.1103/PhysRevA.102.033502 (2020).

28. Bai, Y., Yan, J., Lv, H. & Yang, Y. Plasmonic vortices: a review. *J. Opt.* **24,** 84004; 10.1088/2040-8986/ac7d5f (2022).

29. Chen, W., Abeysinghe, D. C., Nelson, R. L. & Zhan, Q. Experimental confirmation of miniature spiral plasmonic lens as a circular polarization analyzer. *Nano letters* **10,** 2075–2079; 10.1021/nl100340w (2010).

30. Carlon Zambon, N. *et al.* Optically controlling the emission chirality of microlasers. *Nature Photon* **13,** 283–288; 10.1038/s41566-019-0380-z (2019).

31. He, C. *et al.* Nonlinear Boost of Optical Angular Momentum Selectivity by Hybrid Nanolaser Circuits. *Nano letters* **24,** 1784–1791; 10.1021/acs.nanolett.3c04830 (2024).

32. Zhao, X. *et al.* Spin-Orbit-Locking Chiral Bound States in the Continuum. *Physical review letters* **133,** 36201; 10.1103/PhysRevLett.133.036201 (2024).

33. Hsu, C. W., Zhen, B., Stone, A. D., Joannopoulos, J. D. & Soljačić, M. Bound states in the continuum. *Nat Rev Mater* **1**; 10.1038/natrevmats.2016.48 (2016).

34. Koshelev, K., Lepeshov, S., Liu, M., Bogdanov, A. & Kivshar, Y. Asymmetric Metasurfaces with High-Q Resonances Governed by Bound States in the Continuum. *Physical review letters* **121,** 193903; 10.1103/PhysRevLett.121.193903 (2018).

35. Koshelev, K., Favraud, G., Bogdanov, A., Kivshar, Y. & Fratalocchi, A. Nonradiating photonics with resonant dielectric nanostructures. *Nanophotonics (Berlin, Germany)* **8,** 725–745; 10.1515/nanoph-2019-0024 (2019).

36. Azzam, S. I. & Kildishev, A. V. Photonic Bound States in the Continuum: From Basics to Applications. *Advanced Optical Materials* **9**; 10.1002/adom.202001469 (2021).

37. Heimig, C. *et al.* Chiral nonlinear polaritonics with van der Waals metasurfaces. *Science advances* **12,** eaeb5631; 10.1126/sciadv.aeb5631 (2026).

38. Lee, I.-H. *et al.* Image polaritons in boron nitride for extreme polariton confinement with low losses. *Nature communications* **11,** 3649; 10.1038/s41467-020-17424-w (2020).

39. Caldwell, J. D. *et al.* Sub-diffractional volume-confined polaritons in the natural hyperbolic material hexagonal boron nitride. *Nature communications* **5,** 5221; 10.1038/ncomms6221 (2014).

40. Keilmann, F. & Hillenbrand, R. Near-field microscopy by elastic light scattering from a tip. *Philosophical transactions. Series A, Mathematical, physical, and engineering sciences* **362,** 787–805; 10.1098/rsta.2003.1347 (2004).

41. Schnell, M. *et al.* Controlling the near-field oscillations of loaded plasmonic nanoantennas. *Nature Photon* **3,** 287–291; 10.1038/nphoton.2009.46 (2009).

42. Hillenbrand, R., Abate, Y., Liu, M., Chen, X. & Basov, D. N. Visible-to-THz near-field nanoscopy. *Nat Rev Mater* **10,** 285–310; 10.1038/s41578-024-00761-3 (2025).

43. Baù, E. *et al.* Tunable polaritonic topologies generated by non-local photonic modes. *Nat. Nanotechnol.*; 10.1038/s41565-026-02174-5 (2026).

44. Wang, W., Srivastava, Y. K., Tan, T. C., Wang, Z. & Singh, R. Brillouin zone folding driven bound states in the continuum. *Nature communications* **14,** 2811; 10.1038/s41467-023-38367-y (2023).

45. Murai, S. *et al.* Engineering Bound States in the Continuum at Telecom Wavelengths with Non-Bravais Lattices. *Laser & Photonics Reviews* **16**; 10.1002/lpor.202100661 (2022).

46. Tsesses, S. *et al.* Optical skyrmion lattice in evanescent electromagnetic fields. *Science (New York, N.Y.)* **361,** 993–996; 10.1126/science.aau0227 (2018).

47. Oliveira, M. de *et al.* Radially and Azimuthally Pure Vortex Beams from Phase-Amplitude Metasurfaces. *ACS Photonics* **10,** 290–297; 10.1021/acsphotonics.2c01697 (2023).

48. Richter, F. U. *et al.* Gradient High-Q Dielectric Metasurfaces for Broadband Sensing and Control of Vibrational Light-Matter Coupling. *Advanced materials (Deerfield Beach, Fla.)* **36,** e2314279; 10.1002/adma.202314279 (2024).

49. Aigner, A., Weber, T., Wester, A., Maier, S. A. & Tittl, A. Continuous spectral and coupling-strength encoding with dual-gradient metasurfaces. *Nature nanotechnology* **19,** 1804–1812; 10.1038/s41565-024-01767-2 (2024).

50. Baù, E. *et al.* Spatially Encoded Polaritonic Ultra-Strong Coupling in Gradient Metasurfaces with Epsilon-Near-Zero Modes. *Advanced materials (Deerfield Beach, Fla.)* **38,** e10402; 10.1002/adma.202510402 (2026).

51. Mangold, F. *et al.* Phonon-polaritonic skyrmions: Transition from bubble- to Néel-type; 10.48550/arXiv.2603.22176 (2026).

52. Gölz, T. *et al.* Revealing Mode Formation in Quasi-Bound States in the Continuum Metasurfaces via Near-Field Optical Microscopy. *Advanced materials (Deerfield Beach, Fla.)* **36,** e2405978; 10.1002/adma.202405978 (2024).

53. Yang, Y. *et al.* Deuterogenic Plasmonic Vortices. *Nano letters* **20,** 6774–6779; 10.1021/acs.nanolett.0c02699 (2020).

54. Wang, J. *et al.* Orbital angular momentum and beyond in free-space optical communications. *Nanophotonics (Berlin, Germany)* **11,** 645–680; 10.1515/nanoph-2021-0527 (2022).

55. Tsesses, S., Cohen, K., Ostrovsky, E., Gjonaj, B. & Bartal, G. Spin-Orbit Interaction of Light in Plasmonic Lattices. *Nano letters* **19,** 4010–4016; 10.1021/acs.nanolett.9b01343 (2019).

56. Engay, E. *et al.* Transverse optical gradient force in untethered rotating metaspinners. *Light, science & applications* **14,** 38; 10.1038/s41377-024-01720-x (2025).

57. Tittl, A. *et al.* Imaging-based molecular barcoding with pixelated dielectric metasurfaces. *Science (New York, N.Y.)* **360,** 1105–1109; 10.1126/science.aas9768 (2018).

58. Miao, P. *et al.* Orbital angular momentum microlaser. *Science (New York, N.Y.)* **353,** 464–467; 10.1126/science.aaf8533 (2016).

59. Cao, T. *et al.* Tuneable Thermal Emission Using Chalcogenide Metasurface. *Advanced Optical Materials* **6**; 10.1002/adom.201800169 (2018).

60. Zhang, Y. *et al.* Polarization vortices of thermal emission. *Science advances* **11,** eadx6252; 10.1126/sciadv.adx6252 (2025).

61. Koshelev, K. *et al.* Nonlinear Metasurfaces Governed by Bound States in the Continuum. *ACS Photonics* **6,** 1639–1644; 10.1021/acsphotonics.9b00700 (2019).

62. Kim, S. *et al.* Chiral electroluminescence from thin-film perovskite metacavities. *Science advances* **9,** eadh0414; 10.1126/sciadv.adh0414 (2023).

63. Vicentini, E. *et al.* Pseudoheterodyne interferometry for multicolor near-field imaging. *Optics express* **31,** 22308–22322; 10.1364/OE.492213 (2023).

**Supplementary Information: Orbital Angular Momentum Locking via Bound States in the Continuum**

Enrico Baù[4], Connor Heimig[1,4], Jonas Biechteler[1,4], Lina Rohrer[1,4], Michael Hirler[1,4], Haoran Ren[2], Stefan A. Maier[2,3], Alexander A. Antonov[1,4], and Andreas Tittl*[,4,1]

[1]Institute of Photonics, Hamburg University of Technology, 21073 Hamburg, Germany.

[2]School of Physics and Astronomy, Monash University, Clayton, Victoria 3800, Australia.

[3]Department of Physics, Imperial College London, London SW7 2AZ, United Kingdom.

[4]Chair in Hybrid Nanosystems, Nano-Institute Munich, Department of Physics, LMU Munich, Germany.

*Corresponding Author: andreas.tittl@tuhh.de

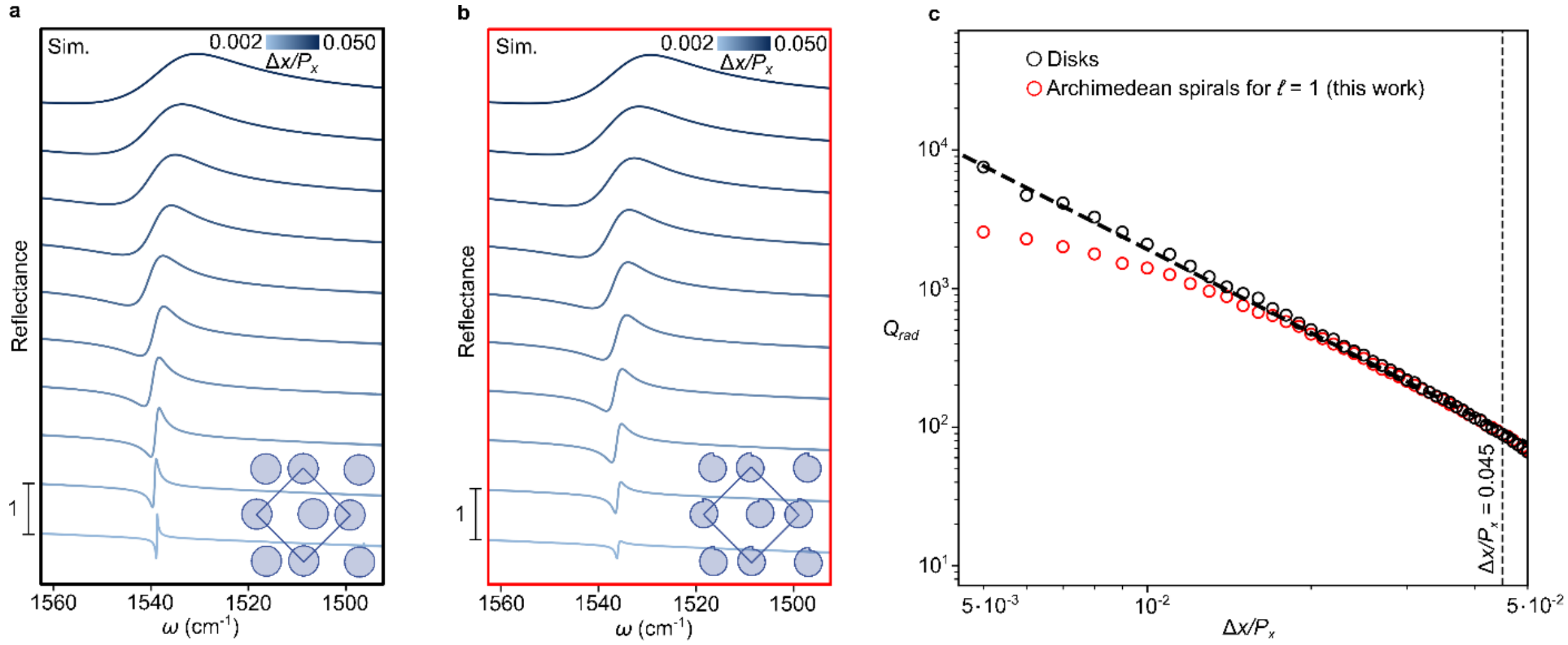

**Figure S1. Simulated spectra of PV-generating unit cell. a**, **b** Simulated reflectance spectra for various asymmetries $\Delta x/P_x$ between 0.002 and 0.05 for the qBIC without and with perturbation for OAM shown in **Fig. 2a** respectively. **c** Radiative $Q$-factor fitted to the spectra shown in (**a**) (black) and (**b**) (red). The black dashed line shows a logarithmic fit in the case of no perturbation for OAM, showing that $Q \propto (\Delta x/P_x)^{-2}$, a defining feature of qBICs[1]. The asymmetry used in all experiments in our work was $\Delta x/P_x$= 0.045.

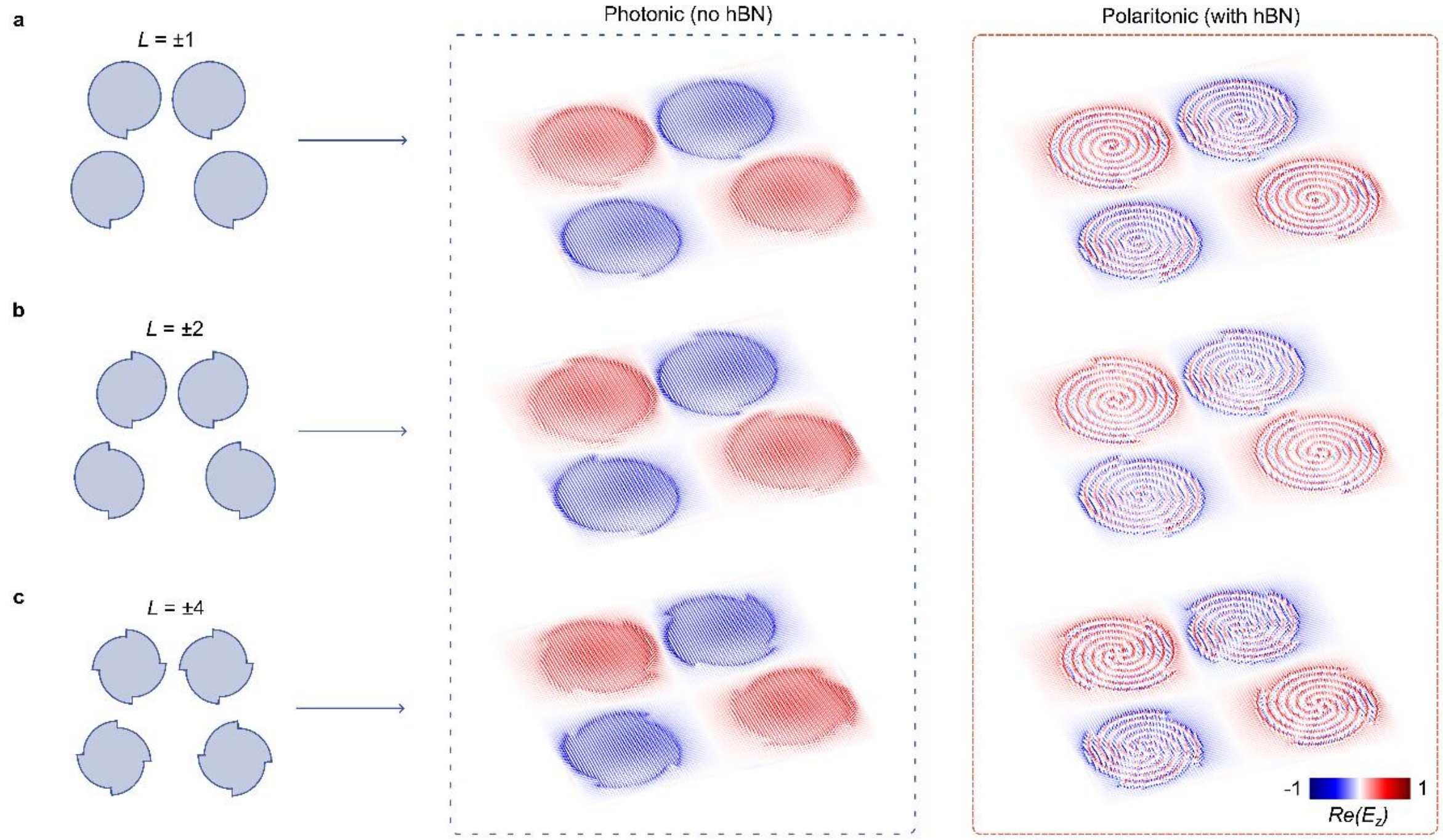

**Figure S2. Photonic and polaritonic near fields.** Simulated electric field for both photonic (no hBN) and polaritonic (covered by hBN) structures with $L = \pm 1$ (**a**), $L = \pm 2$ (**b**) and $L = \pm 4$ (**c**), from top to bottom. Fields distributions are shown 5 nm above the hBN layer.

**Supplementary Note 1**

To model the complex permittivity $\varepsilon_{||}$ of hBN, we can use a Lorentz oscillator of the following form:

$$\varepsilon_j(\omega) = \varepsilon_{j,\infty}(1 + \frac{\omega_{LO,j}^2 - \omega_{TO,j}^2}{\omega_{TO,j}^2 - \omega^2 - i\omega\gamma_j}) \quad (1)$$

where $\varepsilon_{j,\infty}$ denotes the permittivity at high frequencies, $\omega_{TO,j}$ and $\omega_{LO,j}$ are the frequencies of the transverse and longitudinal optical phonon respectively and $\gamma_j$ the damping constant, with the index $j$ denoting either the in-plane or out-of-plane components. For all simulations and calculations conducted in this work, the in-plane permittivity can be modeled using $\varepsilon_{||,\infty} = 4.9$, $\omega_{LO,||} = 1614\ \mathrm{cm}^{-1}$ cm-1, $\omega_{TO,||} = 1360\ \mathrm{cm}^{-1}$, and $\gamma_{||} = 7\ \mathrm{cm}^{-1}$ and the out-of-plane permittivity using $\varepsilon_{z,\infty} = 2.95$, $\omega_{LO,z} = 825\ \mathrm{cm}^{-1}$, $\omega_{TO,z} = 760\ \mathrm{cm}^{-1}$, and $\gamma_z = 2\ \mathrm{cm}^{-1}$. The values were taken from Ref.[2]

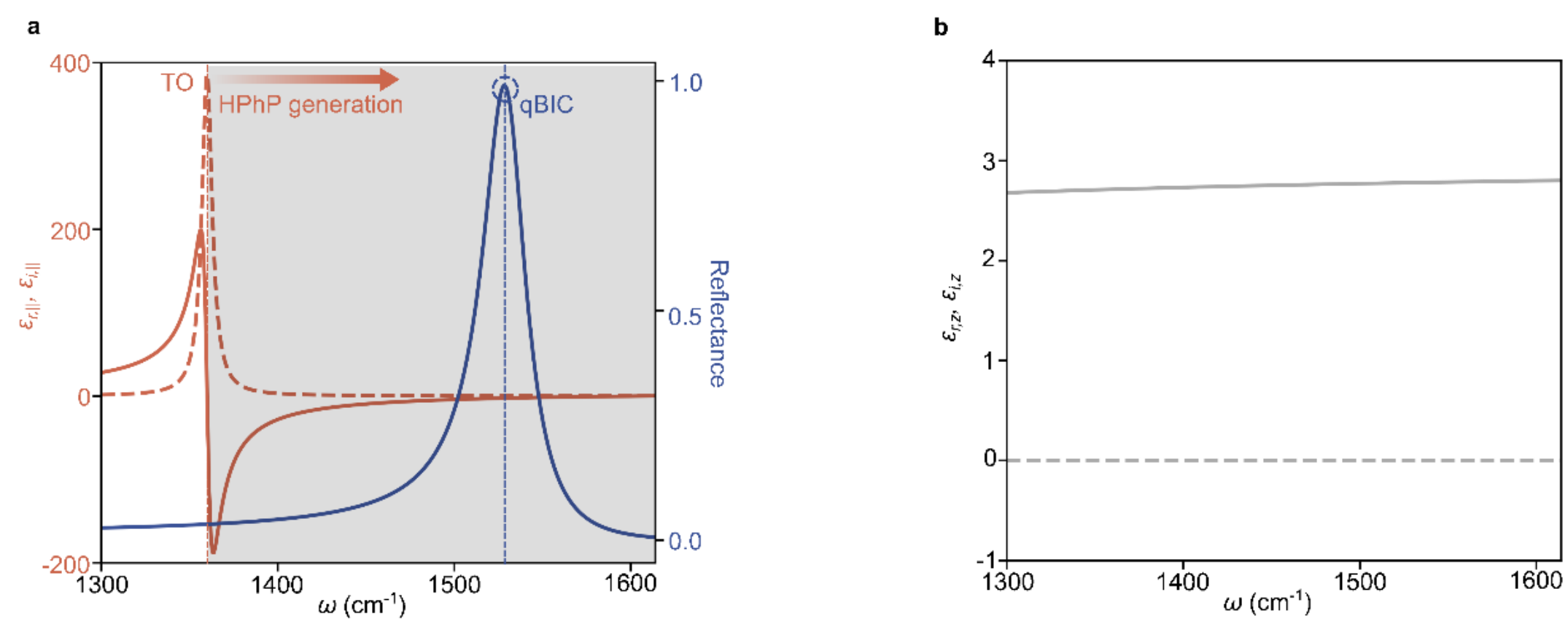

**Figure S3. Permittivity of hBN. a** Real part $\varepsilon_{r,||}$ (full orange curve) and imaginary part $\varepsilon_{i,||}$ (dashed orange curve) of the in-plane permittivity of hBN and simulated reflectance spectrum (blue curves) of the metasurface shown in **Fig. 1a**. The arrow and dark shaded area show the in-plane Reststrahlen-band of hBN where HPhPs can be excited. **b** Real part $\varepsilon_{r,z}$ (full grey curve) and imaginary part $\varepsilon_{i,z}$ (dashed grey curve) of the out-of-plane permittivity of hBN.

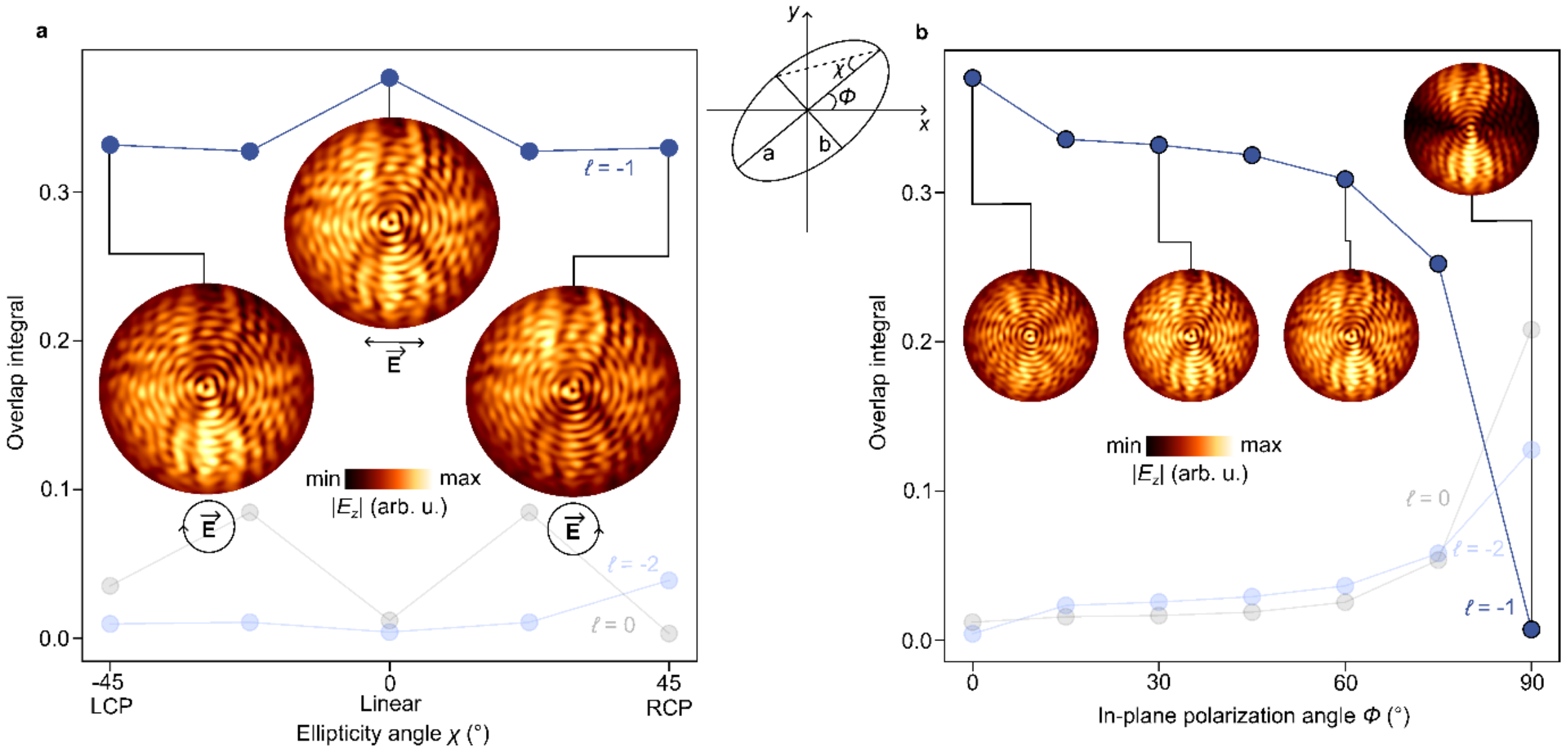

**Figure S4. Robustness of PVs to input polarization state. a** Simulated overlap integral of PVs generated from the structure shown in **Fig. 1a** ($\ell < 0$), excited at $\omega_{PV} = 1535\ \text{cm}^{-1}$ for varying $\chi$ ranging from -45° (LCP) to +45° (RCP). Curves show calculated overlap integrals for $\ell = 0$ (light gray), $\ell = -1$ (dark blue) and $\ell = -2$ (light blue). The insets show the calculated amplitude of the out-of-plane electric field $|E_z|$ of selected PVs. Simulations were conducted for $h_{hBN} = 50$ nm. **b** Simulated overlap integral for linear polarization with various $\Phi$ for $\ell = 0$ (light gray), $\ell = -1$ (dark blue) and $\ell = -2$ (light blue). The insets show the calculated amplitude of the out-of-plane electric field $|E_z|$ of selected PVs. The sketch between (**a**) and (**b**) shows the polarization ellipse with major axis $a$, minor axis $b$, orientation angle of the polarization $\Phi$ and ellipticity angle $\chi$. For large polarization orientation angles $|\Phi| > 70°$, the qBIC does not drive the PV and HPhPs are directly launched through the linearly polarized excitation instead of the photonic mode.

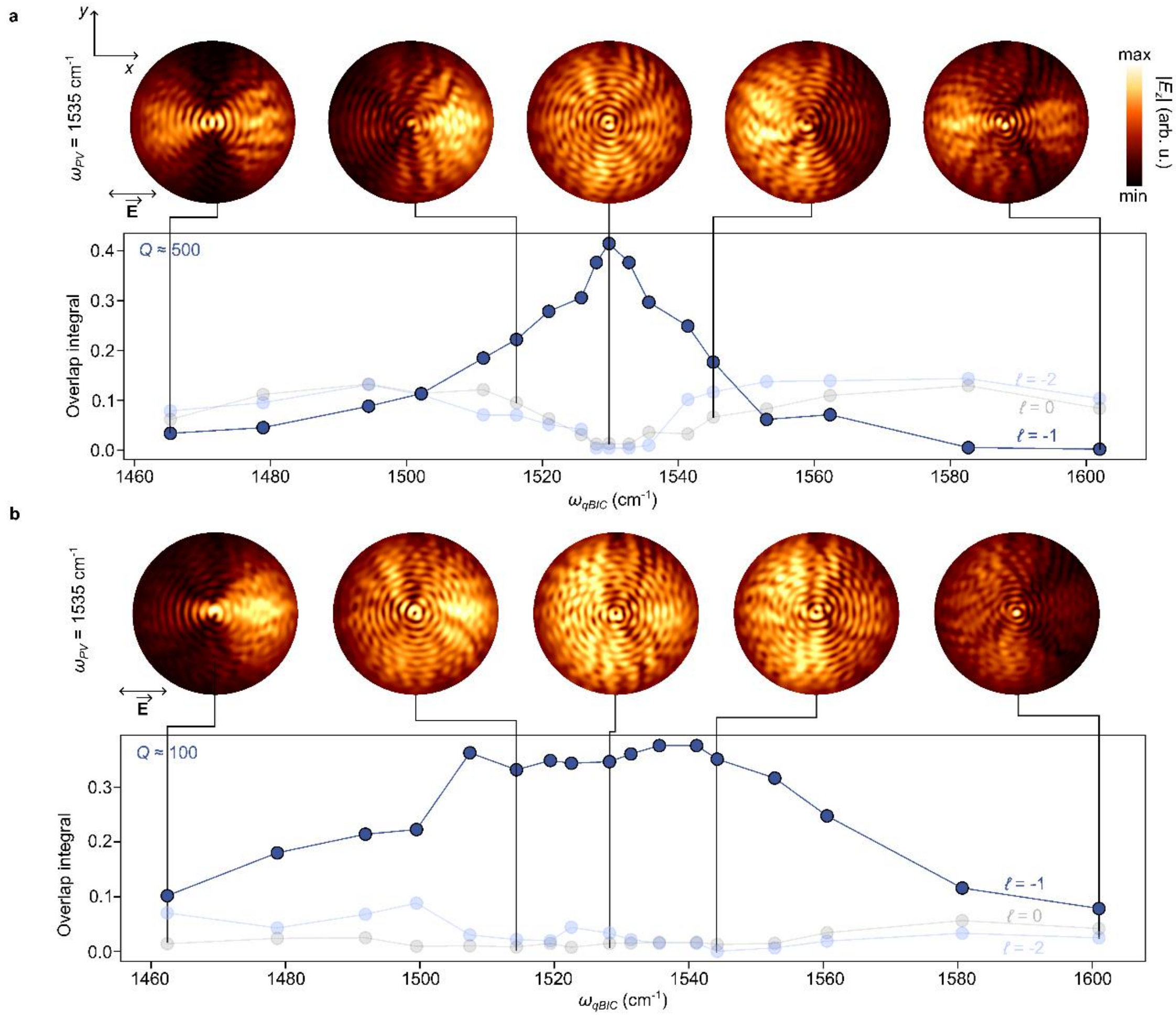

**Figure S5. Locking of OAM for various $Q$-factors and input polarizations. a**, **b** Simulated overlap integral of PVs generated from the structure shown in **Fig. 1a** ($\ell < 0$), excited at $\omega_{PV} = 1535\ \mathrm{cm}^{-1}$ with *x*-polarized light for varying qBIC spectral positions $\omega_{qBIC}$ and $Q$-factors of $Q \approx 500$ (**a**) and $Q \approx 100$ (**b**). Curves show calculated overlap integrals for $\ell = 0$ (light gray), $\ell = -1$ (dark blue) and $\ell = -2$ (light blue). The insets show the calculated amplitude of the out-of-plane electric field $|E_z|$ of selected PVs. Simulations were conducted for $h_{hBN} = 50$ nm and pitches ranging from $P_x = 5150 - 5950$ nm, $P_y = 4635 - 5355$ nm. The distance $m\lambda_{\mathrm{HPhP}}$ with $m = 1$ is kept constant for all simulations.

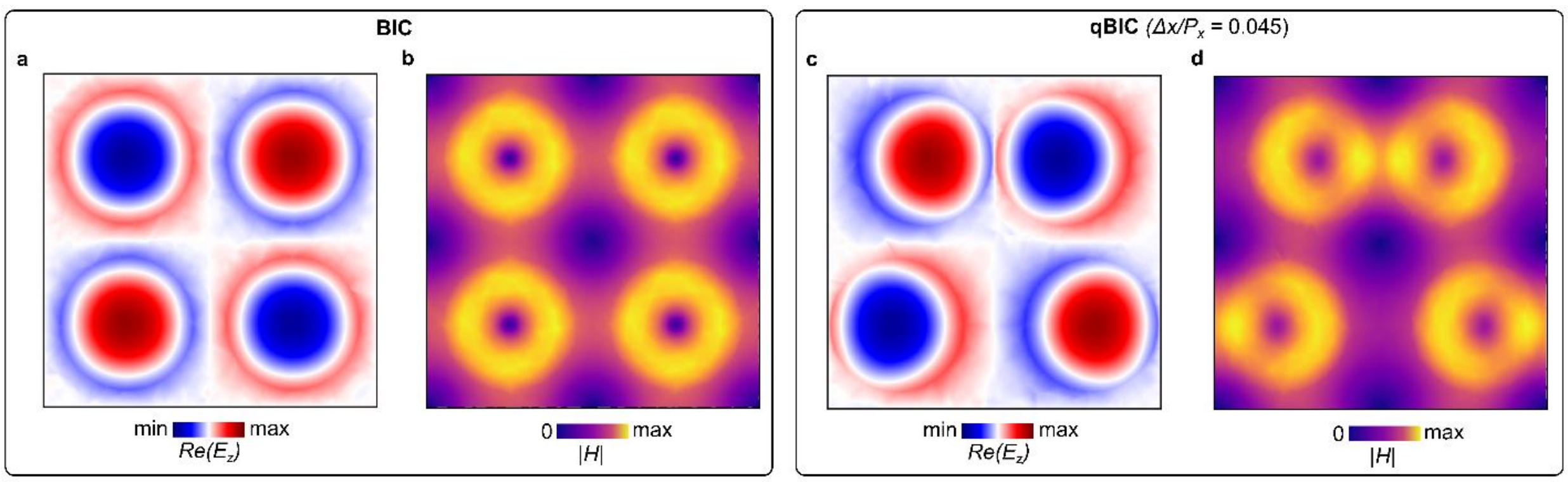

**Figure S6. Eigenmode electric and magnetic fields.** Real part of the out-of-plane electric field $E_z$ and in-plane magnetic field $|H|$ of the BIC (**a**, **b**) and qBIC (**c**, **d**) shown in **Fig. 2a** (left panel). Fields were extracted at a cross-section located at the mid-plane of the resonator along the z-direction.

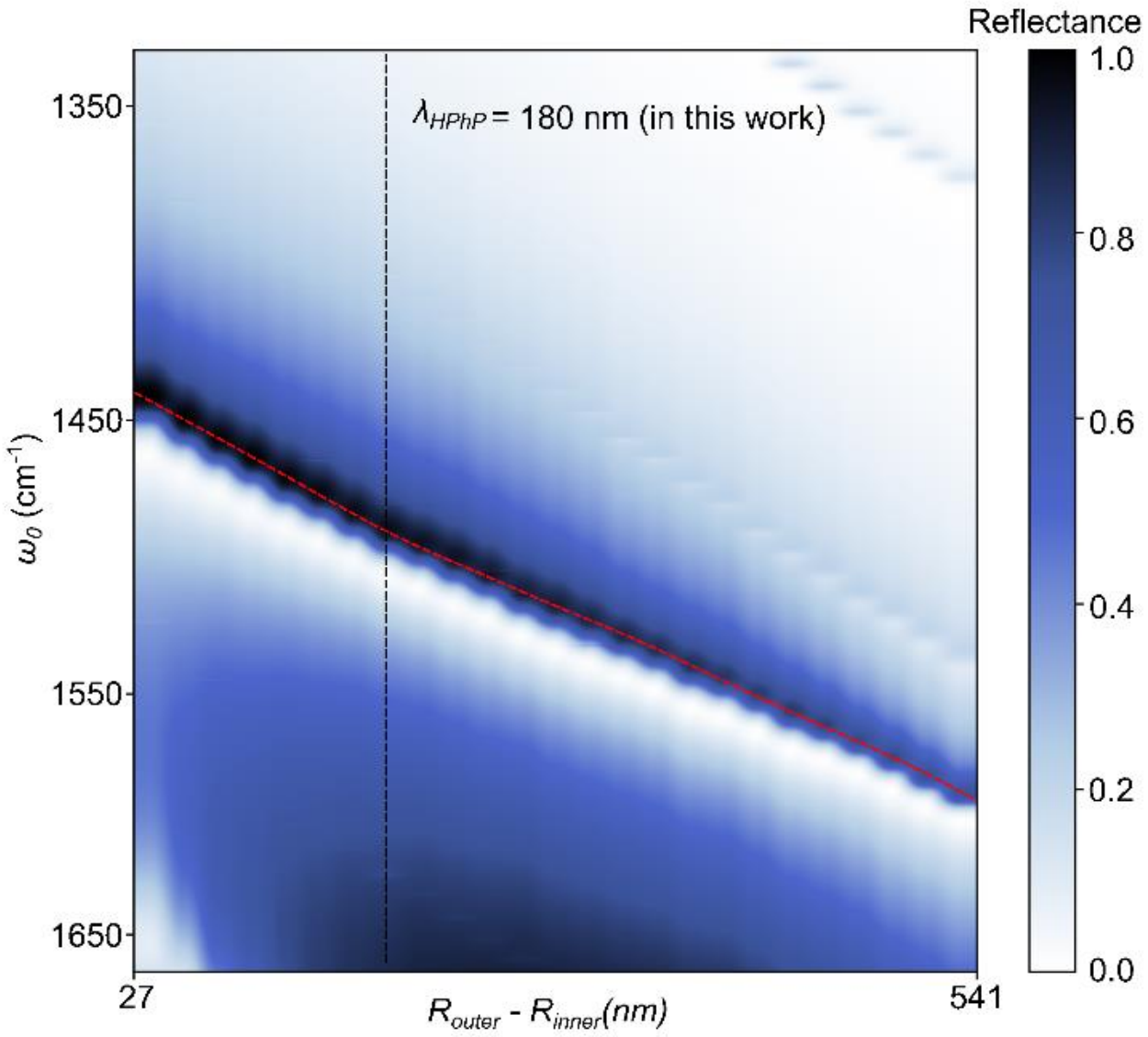

**Figure S7. Reflectance spectra for various Archimedean spirals.** Simulated reflectance spectra for various distances between outer and inner radius of the Archimedian spiral resonators $R_{outer} - R_{inner}$. High modulation (Reflectance>0.5) can be still achieved even for $R_{outer} - R_{inner} >$ 500 nm, which are typically considered to be relatively large for HPhPs[2]. Black dashed line marks the $\lambda_{HPhP} = 180$ nm used for simulations in our work, while red dashed line marks the qBIC.

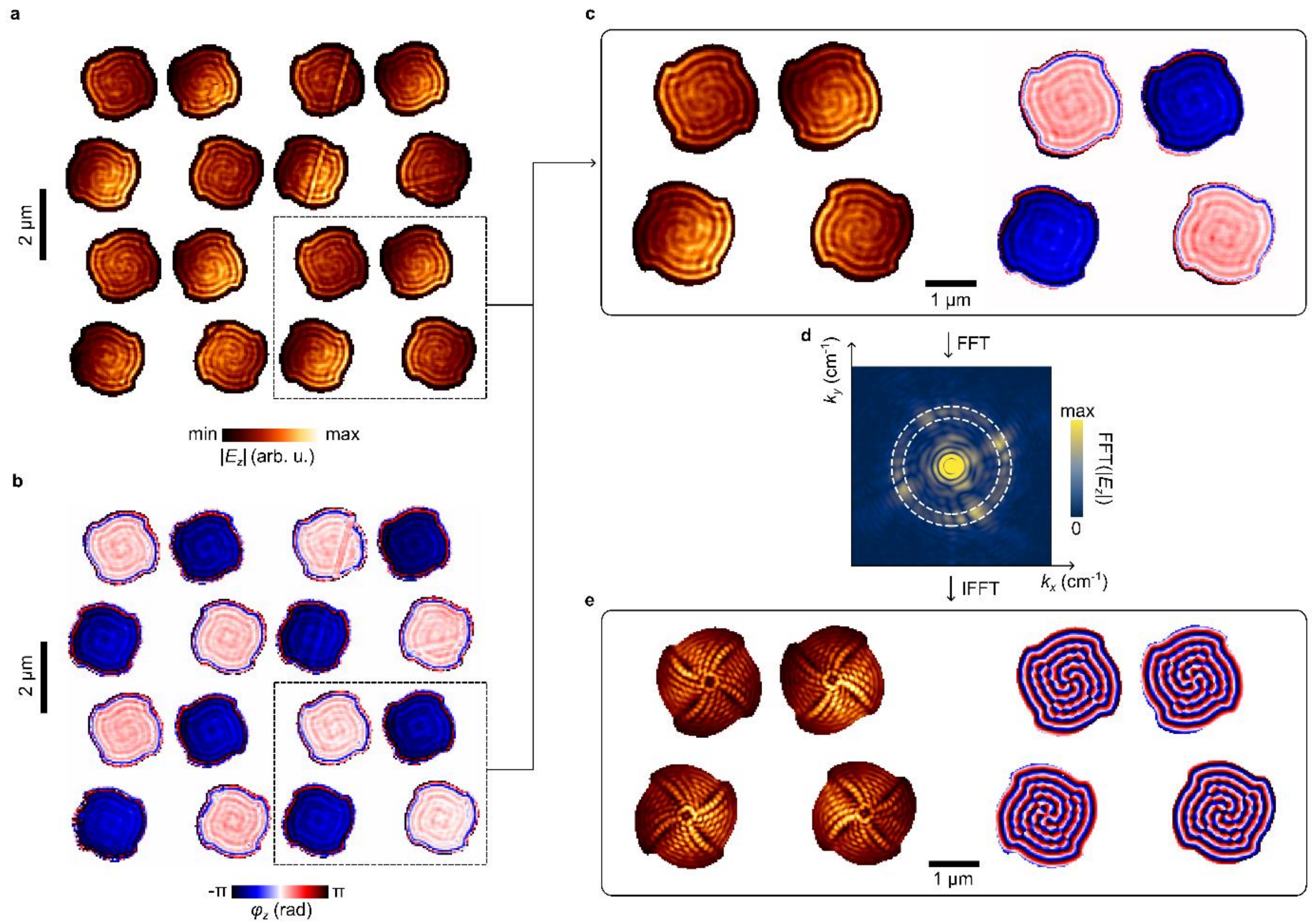

**Figure S8. Fourier-filtering of experimentally measured PVs.** From the out-of-plane amplitude $|E_z|$ (**a**) and phase $\varphi_z$ measured with s-SNOM (**b**), a single unit cell is extracted (**c**) and Fourier-transformed (**d**) to filter-out artefacts and higher frequency components. A suitable region is chosen based on the calculated HPhP dispersion (**Fig. 2g**) and all frequencies below and above 10% of the calculated HPhP momentum are discarded. **e** An Inverse Fourier Transform yields the filtered images used for further analysis in **Fig. 3** and **Fig. 4**.

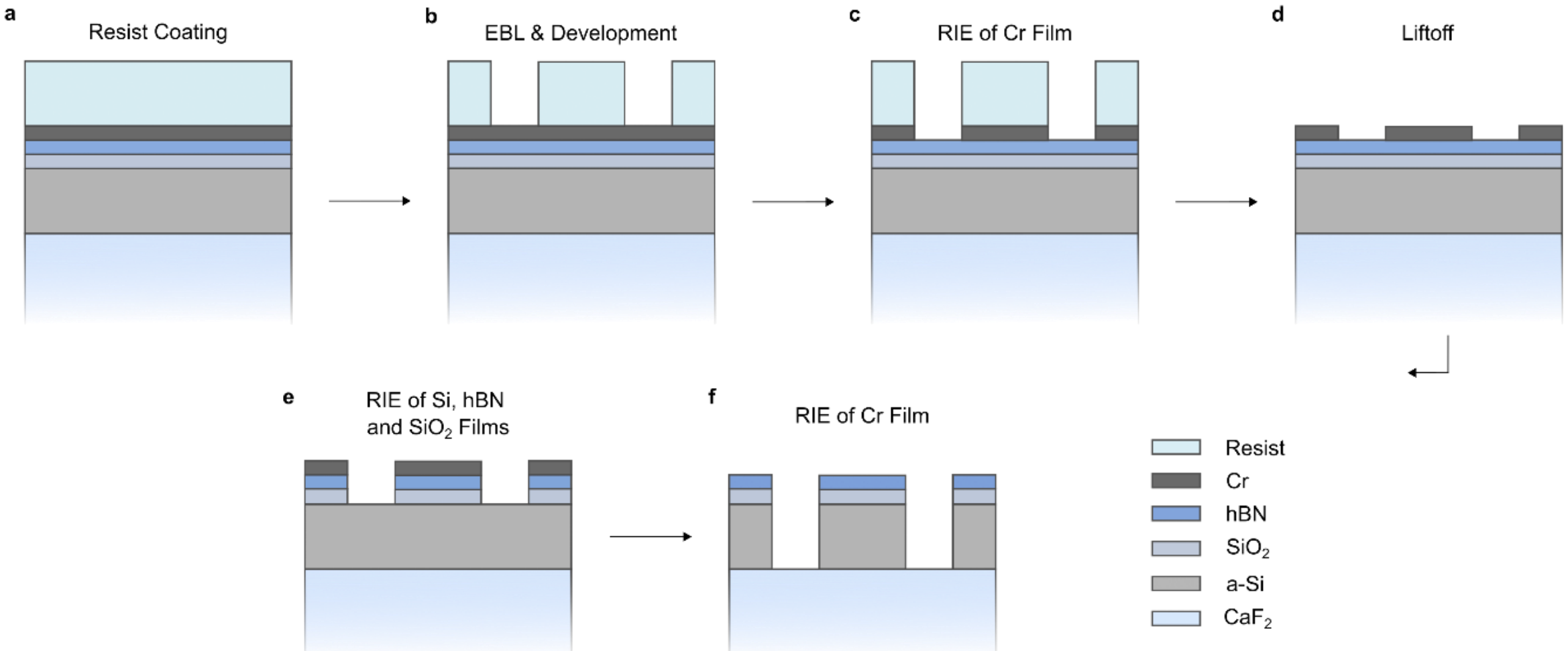

**Figure S9. Patterning process for the OAM qBIC gradient metasurfaces. a** The patterning sequence starts with spin-coating CSAR resist onto the multilayer stack ($CaF_2$/Si ($h_{Si} = 1450$ nm) /$SiO_2$ ($h_{SiO2} = 50$ nm) /hBN ($h_{hBN} = 50 - 100$ nm) /Cr ($h_{Cr} = 100$ nm)), followed by electron beam lithography to define the inverse pattern of the hexagonal structures. **b** After development, the unexposed regions of the resist remains in place and acts as a soft mask. **c** This mask is then used to transfer the pattern into the Cr layer via RIE. **d** Remaining resist is lifted off. **e** The patterned Cr subsequently serves as a durable hard mask during reactive ion etching, which transfers the design into the underlying hBN, $SiO_2$, and Si layers. **f** The process concludes with the removal of the residual Cr yielding the final device structures.

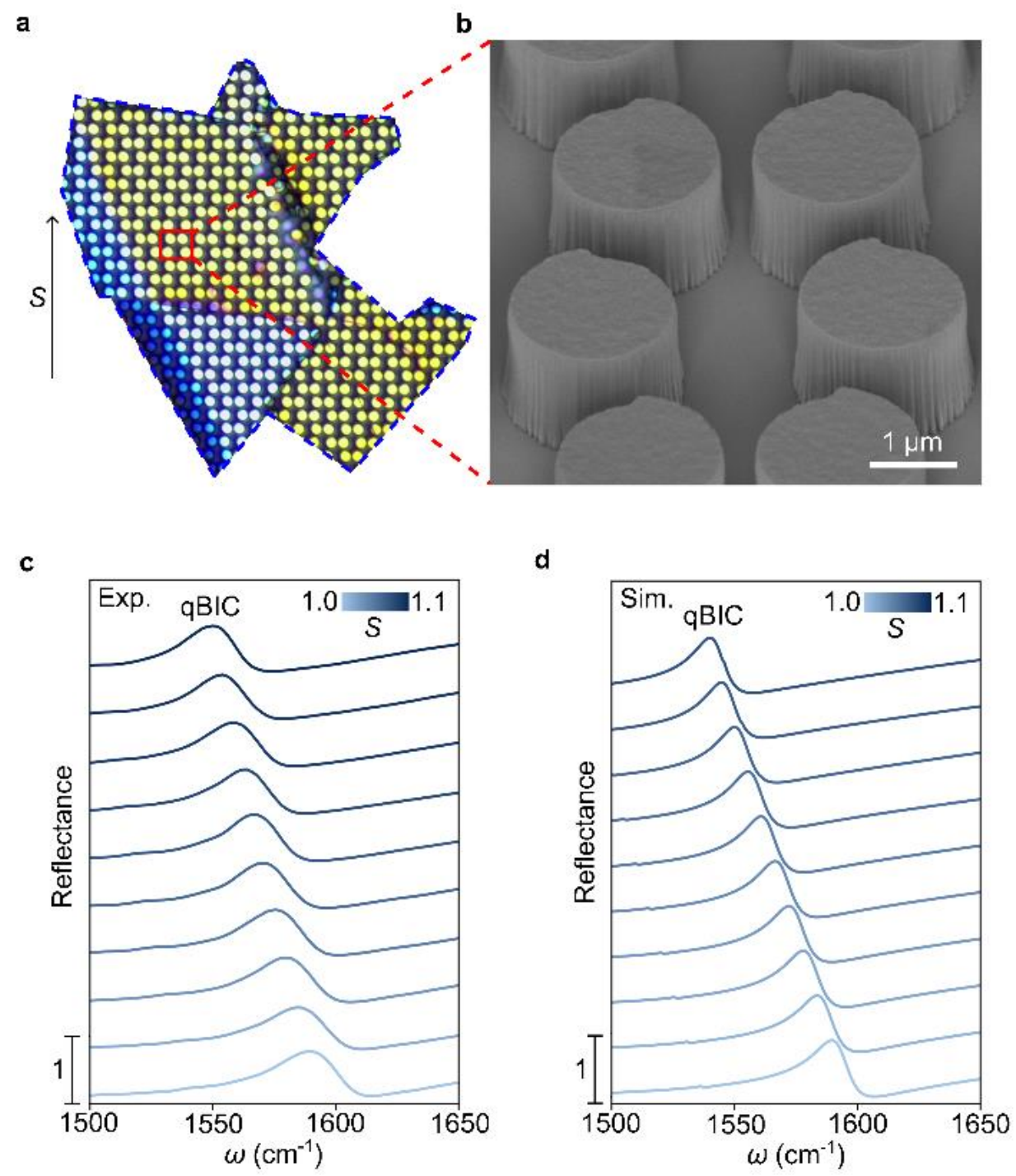

**Figure S10. Fabricated OAM-generating qBIC metasurface. a** Optical image of a metasurface for the generation of PVs with $\ell = 1$. The array was fabricated as a spectral gradient with varying scaling factor $S$ along one spatial axis. hBN flakes of various thicknesses are covering the array. **b** SEM image of a single unit cell for $\ell = 1$. **c** Experimental and **d** simulated reflectance spectra for the metasurface shown in (**a**), for various scaling factors $S$, showing excellent agreement between simulation and experiment.

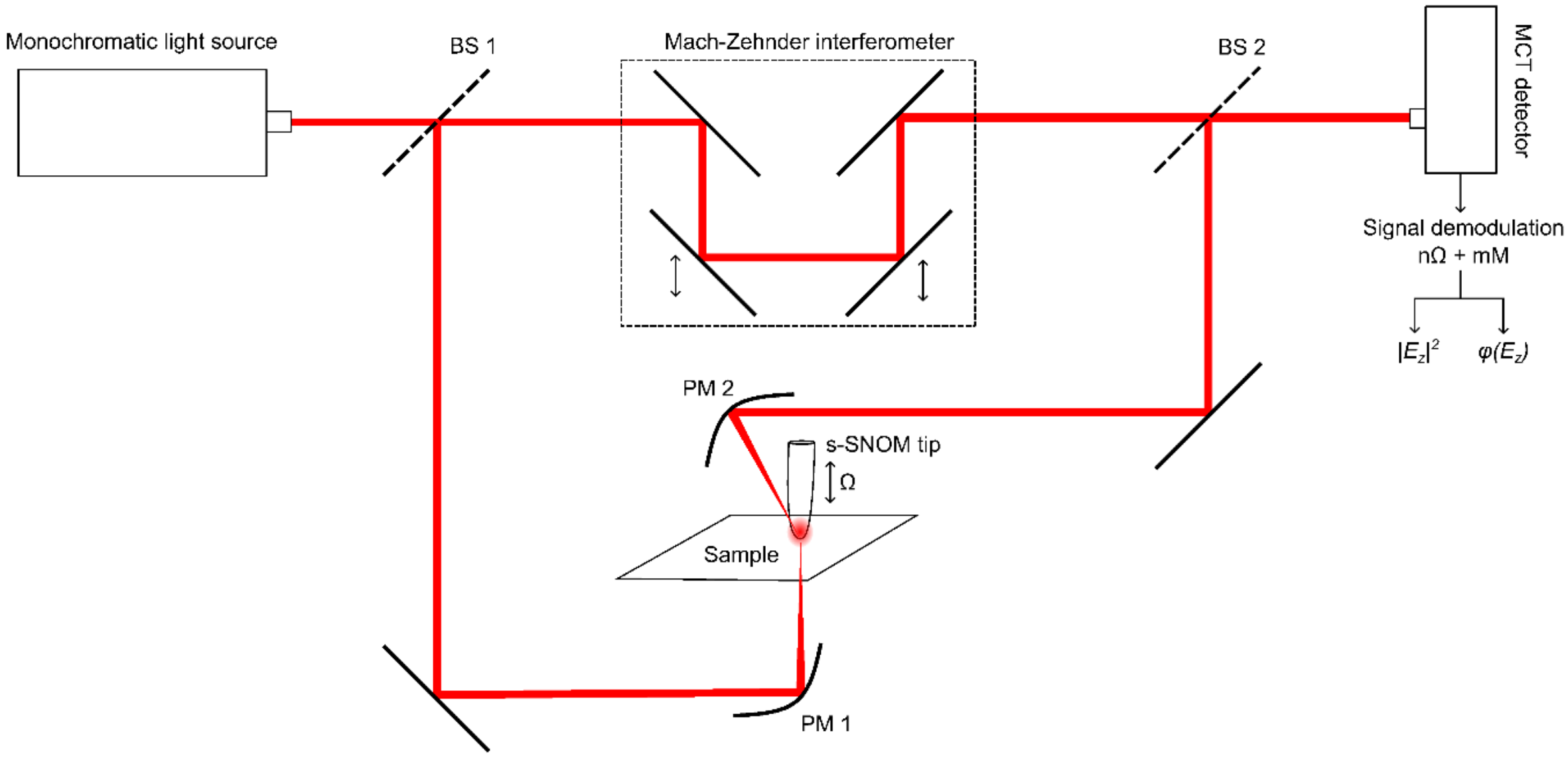

**Figure S11. Transmission-mode phase-resolved near-field imaging.** To measure the near-field optical response of our metasurfaces, we employ a monochromatic light source which passes through a beamsplitter (BS1) and is divided into two parts. One part is focused onto the sample and the tapping s-SNOM tip via a parabolic mirror (PM1) and the backscattered light is collected via a second parabolic mirror (PM2). The other part passes through a delay stage with two vibrating mirrors. Both beams are overlapped at a second beamsplitter (BS2) and the interference is measured with a liquid nitrogen-cooled mercury cadmium telluride detector. Due to the strong polarizability in the direction of the shaft, the tip mostly detects out-of-plane components of the electric field. Both amplitude $|E_z|$ and phase $\varphi_z$ of the out-of-plane electric field can be decoupled and independently characterized through pseudo-heterodyne detection. Exploiting the tip oscillation at a frequency of around $\Omega \approx 250$ kHz, the signal is demodulated at harmonics $n_{harm}\Omega$ to suppress unwanted background ($n_{harm} > 2$ for all measurements).

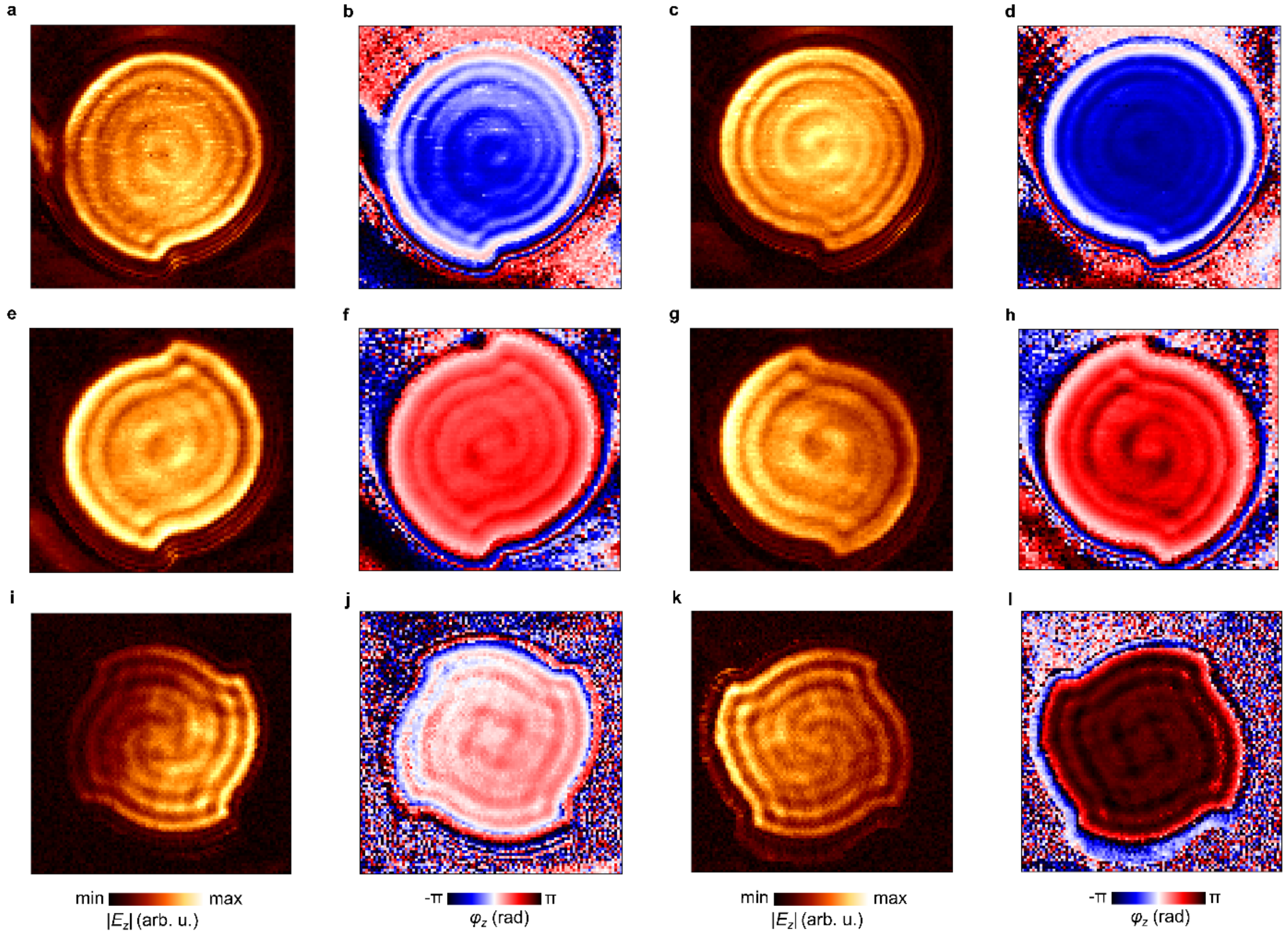

**Figure S12. Unfiltered near-field images of PVs. a, b** Out-of-plane optical amplitude $|E_z|$ and phase $\varphi_z$ of a single resonator generating a PV with an OAM of $\ell = -1$. Same for **c, d** $\ell = +1$, **e, f** $\ell = -2$, **g, h** $\ell = +2$, **i, j** $\ell = -4$ and **k, l** $\ell = +4$.

## References

1. Koshelev, K., Lepeshov, S., Liu, M., Bogdanov, A. & Kivshar, Y. Asymmetric Metasurfaces with High-Q Resonances Governed by Bound States in the Continuum. *Physical review letters* **121,** 193903; 10.1103/PhysRevLett.121.193903 (2018).

2. Caldwell, J. D. *et al.* Sub-diffractional volume-confined polaritons in the natural hyperbolic material hexagonal boron nitride. *Nature communications* **5,** 5221; 10.1038/ncomms6221 (2014).